\documentclass[pre,twocolumn,showpacs,preprintnumbers,amsmath,amssymb]{revtex4}

\usepackage{graphicx} % Include figure files
\usepackage{dcolumn}  % Align table columns on decimal point
\usepackage{bm}       % bold math

\newcommand{\beq}[1][\theequation]{\begin{equation}\label{eq:#1}}
\newcommand{\eeq}{\end{equation}}

\newcommand{\Jbf}{\mathbf{J}}
\newcommand{\Fbf}{\mathbf{F}}
\newcommand{\nbf}{\mathbf{n}}

\newcommand{\ubf}{\mathbf{u}}

\newcommand{\nablabf}{\boldsymbol{\nabla}}

\newcommand{\id}{\mathrm{d}} % integral d symbol
\newcommand{\ext}{\mathrm{ext}}

\newcommand{\Comsol}{\textsc{Comsol}}

\begin{document}

%\preprint{APS/123-QED}

\title{Strongly nonlinear dynamics of electrolytes in large ac voltages}

\author{Laurits H\o{}jgaard Olesen,$^1$ Martin Z. Bazant,$^{2,3}$ and Henrik Bruus$^1$}
\affiliation{%
  $^1$Department of Micro and Nanotechnology, Technical University of Denmark, DK-2800 Kgs.~Lyngby, Denmark\\
  $^2$Departments of Chemical Engineering and Mathematics, Massachusetts Institute of Technology, Cambridge, MA 02139,
  USA\\
  $^3$UMR Gulliver ESPCI-CNRS 7083, 10 rue Vauquelin, F-75005 Paris, France
}%

\date{24 August 2009}

%%%%%%%%%%%%%%%%%%%%%%%%%%%%%%%%%%%%%%%%%%%%%%%%%%%%%%%%%%%%%%%%%%%%%%%%%%%%%%%
\begin{abstract}
We study the response of a model micro-electrochemical cell to a large ac
voltage of frequency comparable  to the inverse cell relaxation time. To
bring out the basic physics, we consider the simplest possible model of a
symmetric binary electrolyte confined between parallel-plate blocking
electrodes, ignoring any transverse instability or fluid flow. We analyze
the resulting one-dimensional problem by matched asymptotic expansions in
the limit of thin double layers and extend previous work into the strongly
nonlinear regime, which is characterized by two novel features --
significant salt depletion in the electrolyte near the electrodes and, at
very large voltage, the breakdown of the quasi-equilibrium structure of the
double layers.  The former leads to the prediction of ``ac capacitive
desalination", since there is a time-averaged transfer of salt from the
bulk to the double layers, via oscillating diffusion layers. The latter is
associated with transient diffusion limitation, which drives the formation
and collapse of space-charge layers, even in the absence of any net
Faradaic current through the cell.  We also predict that steric effects of
finite ion sizes (going beyond dilute solution theory) act to suppress the
strongly nonlinear regime in the limit of concentrated electrolytes, ionic
liquids and molten salts. Beyond the model problem, our reduced equations
for thin double layers, based on uniformly valid matched asymptotic
expansions, provide a useful mathematical framework to describe additional
nonlinear responses to large ac voltages, such as Faradaic reactions,
electro-osmotic instabilities, and induced-charge electrokinetic phenomena.
\end{abstract}

\pacs{ }% PACS, the Physics and Astronomy
\maketitle

%%%%%%%%%%%%%%%%%%%%%%%%%%%%%%%%%%%%%%%%%%%%%%%%%%%%%%%%%%%%%%%%%%%%%%%%%%%%%%%
\section{Introduction\label{sec:introduction}}

Time-dependent voltages are applied to electrolytes in many different
fields, and theoretical models to interpret the results have been
developed for over a century~\cite{bazant2004}. Current applications
include energy storage in electrochemical systems
(e.g. supercapacitors~\cite{kotz2000,jang2005,eikerling2005},
high-rate batteries~\cite{dudney1995,wang1996,takami2002}), flow
control in microfluidics (e.g. ac
electroosmotic~\cite{squires2005,ramos1999,ajdari2000,olesen2006,studer2004,urbanski2006, gregersen2007,iceo2004b,levitan2005,cahill2004,brask2006}
and electrothermal~\cite{gonzalez2006,wu2007} flows), particle
handling in colloidal materials
(e.g. dielectrophoresis~\cite{green2000,wong2004}, induced-charge
electrophoresis~\cite{iceo2004a,murtsovkin1996,squires2006,gangwal2008}),
cellular and molecular manipulation in the biological systems
(e.g. electroporation~\cite{weaver1993,weaver1996,lu2005}, cell
trapping~\cite{voldman2006,wu2005,wong2004}, and biomolecular
sensing~\cite{bard1992,chen2000,zhang2004,heien2005,chi2005}).

In many cases, periodic voltages are used to drive alternating
current (ac) to eliminate any net linear response, such as direct
current or electroosmotic flow. The most common application of ac
forcing is in impedance spectroscopy, long used to characterize
electrochemical interfaces~\cite{bard_book}.  The current response to
a small sinusoidal voltage is fitted to an electrical circuit model,where the interface acts an impedance in series with a bulk
resistance~\cite{sluyters1970,macdonald1990,geddes1997}. The characteristic
frequency for double-layer charging is then the inverse ``$RC$ time" of
the equivalent circuit~\cite{bazant2004}. Circuit models are also used
to describe electrochemical response in much more complicated
situations, such as composite porous
electrodes~\cite{jang2005,eikerling2005}, micro-electrode
arrays~\cite{ramos1999,ajdari2000,levitan2005,wu2005,gregersen2007},
and biological tissues~\cite{weaver1993,weaver1996}.

Circuit models can be derived from underlying ion-transport equations
by considering the joint limit of thin double layers ($\epsilon =
\lambda_D/L \ll 1$, where $\lambda_D$ is the Debye--H\"uckel screening
length and $L$ is the geometrical scale) and small voltages ($V \ll kT/e$
where $V$ is the amplitude of the applied voltage and $kT/e$ is the
thermal voltage)~\cite{bazant2004}. From a
mathematical point of view, this can be done systematically starting
from the Poisson--Nernst--Planck equations (PNP) by asymptotic
boundary-layer analysis, which was introduced to electrochemistry in
the 1960s to justify the thin-double-layer
approximation~\cite{grafov1962,chernenko1963,newman1965,macgillivray1968,bazant2005}. The
joint asymptotic limit of thin double layers and large voltages, which is
mathematically more challenging and physically more complex, has also
been analyzed under conditions of steady direct current (dc). At
sufficiently large \emph{steady} dc currents, exceeding diffusion
limitation, a variety of exotic effects arise, such as the expansion
of the double layer into an extended, non-equilibrium space-charge
layer~\cite{smyrl1967,rubinstein1979,chu2005} and electro-hydrodynamic
instability due to second-kind electroosmotic
flows~\cite{rubinstein2000,rubinstein2001,zaltzman2007}. Clearly, such
effects cannot be captured by classical circuit models, but they
continue to be used in dynamical situations, even with large
voltages, for lack of a simple mathematical alternative.

The \emph{transient} electrochemical response to a large dc voltage
(without Faradaic reactions) has only been analyzed quite
recently~\cite{bazant2004,chu2006,kilic2007a,kilic2007b,beunis2008}.
Even in the
limit of thin double layers, the large voltage leads to a number of
new dynamical effects not captured by circuit models. Additional time
scales enter the problem, other than the fundamental $RC$ time scale
(which can be expressed as $\lambda_D L/D$ where $D$ is the ion diffusivity~\cite{bazant2004}). In the
simplest one-dimensional problem with parallel-plate, blocking
electrodes, over-charging of the double-layer ``capacitors" leads to
net adsorption of neutral salt from the bulk, regardless of the
polarity of the diffuse-layer voltage~\cite{bazant2004}. This process
is coupled to slow diffusive relaxation of the bulk concentration (at
the time scale $L^2/D$), which leads to transient concentration
polarization and thus breakdown of Ohm's law for the bulk
``resistor". In higher dimensions, large applied voltages also trigger
surface transport of ions through the double layers~\cite{chu2007},
which completes flux loops driven by bulk concentration gradients in
and out of the double layers~\cite{chu2006}.

At large voltages, another important consideration is the breakdown of dilute-solution theory~\cite{large,large2}, including the Poisson--Boltzmann (PB) model of the double
layer~\cite{kilic2007a}, and, more generally, the PNP equations
from which it is derived~\cite{kilic2007b}. These classical models are
strictly valid only for a dilute solution of point-like ions, but,
even in a very dilute bulk solution, the application of a large
voltage can lead to the crowding of counterions near a highly charged
electrode. Among many possible modifications of the PB model, one must
at least account for the finite sizes of ions and solvent
molecules. This generally leads to the formation of a condensed layer
of crowded ions, anticipated by Stern~\cite{stern1924} and first described 1942 by
Bikerman~\cite{bikerman1942}, whose simple modified PB (MPB) model has
been rederived several times in different
contexts~\cite{dutta1950,bagchi1950,eigen1951,eigen1954,iglic1994,kralj-iglic1996,borukhov1997}. As
perhaps first predicted by Freise in 1952~\cite{freise1952}, the widening of the
condensed layer generally causes the diffuse-layer differential
capacitance to decay at large voltages -- the opposite trend from PB
theory, which allows ions to pile up with exponentially diverging
concentration. This has major implications for the dynamics of
electrolytes at large voltages~\cite{large,large2,kilic2007a,storey2008}
as well as ionic liquids and molten salts~\cite{kornyshev2007,federov2008,federov2008b,oldham2008} (where
crowding dominates in the absence of a solvent).  In electrolytes, for the same
reason, steric constraints also greatly reduce salt adsorption and
surface conduction compared to PB theory by limiting the charge
density of the double layer~\cite{kilic2007a,chu2007}.  All of these
conclusions are independent of the model for steric effects on the
chemical potential of ions in a concentrated
solution~\cite{large,biesheuvel2007} and can be extended to more general
situations, without assuming thin double layers, by deriving modified
PNP (MPNP) equations~\cite{kilic2007b}.

A number of recent developments provide further motivation for our work. In
a recent paper~\cite{beunis2008}, Beunis \emph{et al.} have revisited the
problem of a suddenly applied large dc voltage in a blocking cell and
studied the formation of \emph{transient space-charge layers} at very large
voltage, a possibility predicted in Ref.~\cite{bazant2004} and analyzed
preliminarily in Ref.~\cite{olesen_thesis}.  Two recent papers, by Suh and
Kang~\cite{suh2008,suh2009}, analyze the weakly nonlinear response of an
electrolyte to an ac voltage, which is relevant for many of the
experimental situations described above. By coupling weakly nonlinear,
charge relaxation to fluid flow, novel concentrated-solution effects can
enter theory of induced-charge electrokinetic phenomena in large ac
voltages~\cite{large2}.  In the context of electrodialysis membranes, it is
well known that strongly nonlinear effects are important and can lead to
electro-osmotic instability at the limiting current
~\cite{rubinstein2000,rubinstein2001,zaltzman2007}, but this possibility is
just beginning to be explored experimentally using large ac voltages.
Building on recent observations of salt depletion and electro-convection
near micro/nano-channel junctions \cite{kim2007}, the Rubinstein-Zaltzman
instability has been been demonstrated experimentally by applying
low-frequency ac (square-wave) voltages to confine it to slowly oscillating
boundary layers~\cite{yossifon2008}. This experiment raises interesting
theoretical questions about the periodic breakdown and restoration of the
quasi-equilibrium structure of the double-layer under strong ac forcing,
which are a major focus of this paper.

In this work, we analyze the strongly nonlinear, time-dependent response of an electrolyte or ionic liquid to a large ac voltage, apparently for the first time.  The imposition of a time scale (the ac period) is a significant complication compared to case of a sudden dc voltage, so we focus on the simplest geometry of parallel-plate blocking electrodes and ignore any transverse instability. Following Ref.~\cite{bazant2004}, we analyze the resulting one-dimensional problem starting from the classical PNP equations and derive accurate asymptotic approximations for thin double layers. We also consider the MPNP equations of Ref.~\cite{kilic2007b} to highlight steric effects under ac forcing. Using both PNP and MPNP models, we  study the formation and collapse of transient space-charge layers at large voltages. While Beunis {\it et.al}~\cite{beunis2008} focus on the extreme case where the space-charge layer completely dominates the response at very large voltages (in sufficiently large systems and high salt concentrations), we aim to derive a reduced  model that is uniformly valid for all voltages and all salt concentrations, ranging from dilute electrolytes to concentrated solutions and ionic liquids. In spite of the mathematical complexity of these problems, our goal is to extract generic predictions and useful analytical approximations to aid in interpreting experimental data.

The paper is organized as follows. We begin in Sec.~\ref{sec:governing} by stating the mathematical problem, converting to dimensionless form, and showing full numerical solutions used to test our subsequent analytical approximations. In Sec.~\ref{sec:asymptotics} we briefly go though the asymptotic analysis for double layers in quasi-equilibrium, adapting the results of Ref.~\cite{bazant2004} concerning the transient dynamics, to our case of interest, namely the steady-state response when an ac voltage with frequency around the inverse $RC$ time is applied.
In Secs.~\ref{sec:weakly:nonlinear} and \ref{sec:strongly:nonlinear} we study the dynamic response in the weakly and strongly nonlinear regimes, respectively, and discuss how the circuit model is changed when we go from the weakly to the strongly nonlinear regimes. We also compare the strongly nonlinear asymptotic analysis to the full numerical solution. In Sec.~\ref{sec:asymptotics:beyond} we develop an asymptotic analysis for the case when the double layers are driven \emph{out of quasi-equilibrium} to form bulk space-charge and also compare those results to the full numerical solution. Finally, in Sec.~\ref{sec:conclusion} we summarize and briefly discuss extensions to higher dimensions, Faradaic currents, and nonlinear electroosmotic flows, building on the initial study of Ref.~\cite{olesen_thesis}, and we leave the reader with some open questions.

%%%%%%%%%%%%%%%%%%%%%%%%%%%%%%%%%%%%%%%%%%%%%%%%%%%%%%%%%%%%%%%%%%%%%%%%%%%%%%%
\section{Governing equations\label{sec:governing}}

\subsection{General models\label{sec:general:model}}

In any continuum model, the transport of ions in the electrolyte is
governed by a mass conservation law
\beq[conservation:units]
\partial_t c_i = -\nablabf\cdot\Fbf_i,
\eeq
where $c_i$ is the local concentration of the $i$th ionic species,
$\Fbf_i$ is the flux, and we neglect any bulk reactions in the
electrolyte which could produce or consume ions. Quite generally, in a concentrated solution, the flux can be expressed in terms of
the gradient of the electrochemical potential $\mu_i$ as
\beq[general:flux:units]
\Fbf_i = -\frac{D_i}{kT}\,c_i\nablabf\mu_i + \ubf\,c_i,
\eeq
where the first term describes ion transport by diffusion and
electromigration with diffusivity $D_i$, $k$ is Boltzmann's constant, $T$ is absolute temperature, and the second term describes advection at the mean fluid velocity $\ubf$, as determined by momentum conservation. For a dilute solution, the chemical potential $\mu_i$
takes the ideal form, with contributions from entropy and mean
electrostatic energy,
\beq[chemical:potential:units]
\mu_i = kT\log c_i + z_i e \phi,
\eeq
where $\phi$ is the electrostatic potential,  $z_i$ is the ionic
valence and $e$ the electron
charge. Equation~\eqref{eq:conservation:units} then reduces to the
Nernst--Planck equation
\beq[nernst:planck:units]
\Fbf_i = -D_i\Big(\nablabf c_i + \frac{z_ie}{kT}c_i\nablabf\phi\Big) + \ubf c_i.
\eeq
In the usual mean-field approximation, the electrostatic potential is
self-consistently determined by the charge density $\rho$ through
Poisson's equation
\beq[poisson:units]
-\nablabf\cdot(\varepsilon\nablabf\phi) = \rho  =
\sum_i z_i e c_i
\eeq
where $\varepsilon$ is the electrolyte permittivity, which we take to
be constant. This completes the classical PNP equations, which
underly most of electrochemical transport theory.
As noted above, the characteristic length scale in these equations (the
Debye--H\"uckel screening length) is
\beq
\lambda_D = \sqrt{\frac{\varepsilon kT}{\sum_i c_i^*z_i^2 e^2}}.
\eeq
where $c_i^*$ is the nominal bulk concentration of the $i$th ionic
species.

In the present work we focus on dilute electrolytes for which the
nominal bulk salt concentration is small, seemingly within the range
of applicability of the PNP equations. Even in a very dilute bulk
solution, however, when a large
external bias is placed on the electrodes in the system (only a few
times $kT/e$), ions accumulate at the surface,
and the dilute-solution approximation must break
down~\cite{kilic2007a,large,large2}. Following Kilic \emph{et~al.}~\cite{kilic2007b}, we will solve modified (MPNP) equations based on the
oldest and simplest approach to steric effects of ion crowding of
Bikerman~\cite{bikerman1942}, which corresponds
to the following model for the chemical
potential in a binary $z:z$ electrolyte~\cite{kilic2007b,biesheuvel2007},
\beq[electrochemical:steric:units]
\mu_\pm = kT\log c_\pm \pm z e \phi
- kT\log(1 - c_+a^3 - c_-a^3),
\eeq
where $a$ is an effective molecular
length scale. (For a history of this model and related concentrated-solution theories, see Ref.~\cite{large2}.) The correction term, which can be interpreted as an activity coefficient $f_i = \exp[(\mu_i-\mu_i^\mathrm{ideal})/kT]$, is related to the entropy of the
solvent molecules and imposes a maximum ion concentration $c_\mathrm{max} =
a^{-3}$; it can be derived from the statistical mechanics of
of equal-sized ions and solvent molecules on a cubic lattice of
spacing $a$ in the continuum limit. In equilibrium, $\mu_\pm =$
constant, the ions
effectively obey Fermi--Dirac statistics, rather than classical
Boltzmann statistics, due to the excluded volume
effect~\cite{dutta1950,bagchi1950,eigen1951,eigen1954,iglic1994,kralj-iglic1996,borukhov1997,kornyshev2007}. Although
more sophisticated models for $\mu_i$
exist~\cite{kilic2007b,biesheuvel2007,large2}, this approach at least
qualitatively captures the effects of volume constraints with only one
additional parameter $a$.

For boundary conditions at the (blocking) electrodes, we assume no
electrochemical reactions, so the normal ionic fluxes must vanish
$\nbf\cdot\Fbf_i = 0$. To close the system, we follow many prior
authors~\cite{kornyshev1981,ajdari2000,bonnefont2001,bazant2004,bazant2005,levitan2005,olesen2006}
and allow for a compact (Stern) layer or thin dielectric coating separating the
electrode from the electrolyte with a constant ``surface capacitance"
per unit area $C_S$, which leads to a mixed boundary condition
\beq[stern:bc:units]
 C_S (V_\mathrm{ext}-\phi) + \varepsilon \, \nbf\cdot\nablabf\phi = 0.
\eeq
Here $\nbf$ is a surface normal pointing into the electrolyte, $C_S =
\varepsilon_S/h_S$ can be ascribed to a surface coating of
thickness $h_S$ and dielectric constant $\varepsilon_S$, and
$V_\mathrm{ext}(t)$ is the external potential applied at the electrode.

\begin{figure}
\includegraphics[width=0.25\textwidth]{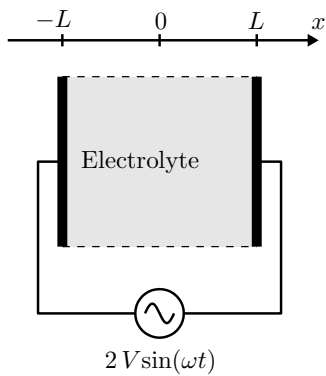}
\caption{Sketch of 1D model problem. The electrolyte is confined
  between parallel-plate blocking electrodes separated by a gap of
  width $2L$, and a harmonic potential of $V_\mathrm{ext}(t)=\pm V\sin(\omega t)$ is
  applied to the left and right electrode, respectively, so the
  overall potential drop across the cell is $2 V \sin(\omega t)$;
  this corresponds to a $4V$ peak-to-peak voltage or $\sqrt{2}V$ rms.}
\label{fig:capacitor:sketch}
\end{figure}

For the present analysis we focus on a symmetric binary electrolyte
with equal diffusivity $D_+=D_-=D$ and valence $z_+=-z_-=z$ for the
two ionic species. Moreover we restrict our attention to the simplest
prototypical microelectrochemical system, consisting of the
electrolyte confined between two parallel, planar, blocking electrodes
at $x=\pm L$, as sketched in Fig.~\ref{fig:capacitor:sketch}. By
symmetry, this rules out any effects of surface
conduction~\cite{chu2006,chu2007} or ac electroosmotic
flow~\cite{ramos1999,ajdari2000,olesen2006} and allows us to focus on
the strongly nonlinear response due to the excessive accumulation of
ions in the screening layers at the electrodes. In summary, the system
is identical to that studied in Refs.~\cite{bazant2004} (PNP) and
~\cite{kilic2007b} (MPNP), except that we apply an ac voltage rather
than a step dc voltage and study the periodic response after all
transients have decayed. We shall see that imposing an external time scale (the ac period)
fundamentally alters the dynamics and complicates the analysis.

%%%%%%%%%%%%%%%%%%%%%%%%%%%%%%%%%%%%%%%%%%%%%%%%%%%%%%%%%%%%%%%%%%%%%%%%%%%%%%%
\subsection{Dimensionless form in one dimension\label{sec:dimensionless}}

We cast the problem into dimensionless form using $L$ as the reference
length scale and the $RC$ relaxation time $\tau = \lambda_D L/D$ as
reference time scale~\cite{bazant2004}, so that time and space are
represented by the dimensionless variables $t' = t/\tau$ and $x' =
x/L$. The potential and ionic concentrations are rescaled as $\phi' =
\phi\, ze/kT$ and $c_\pm' = c_\pm/c^*$, where $kT/e$ is the thermal voltage scale and $c^*$ is
the nominal bulk electrolyte concentration.
%We focus on the thin double layers and the small parameter
%\beq
%\epsilon = \frac{\lambda_D}{L} \ll 1
%\eeq
%is now a numbered equation\ldots

After dropping
the primes from the dimensionless variables, the governing equations
take the form
\begin{align}
-\epsilon^2\partial_x^2\phi &= \frac{1}{2}(c_+-c_-), \label{eq:poisson} \\
\partial_t c_\pm &= -\epsilon\partial_xF_\pm, \label{eq:mass:conservation}
\end{align}
where the fluxes $F_\pm$ are given by
\beq[flux:gradient]
F_\pm = -c_\pm\partial_x\mu_\pm.
\eeq
For a dilute electrolyte the electrochemical potentials reduce to
\beq[chemical:potential]
\mu_\pm = \log c_\pm \pm \phi,
\eeq
and we arrive at the Nernst--Planck equations in dimensionless form
\beq[nernst:planck]
\partial_t c_\pm = \epsilon\partial_x(\partial_x c_\pm \pm c_\pm\partial_x\phi).
\eeq
When steric exclusion is taken into account we get
\beq[electrochemical:steric]
\mu_\pm = \log c_\pm \pm \phi - \log(1-\nu c),
\eeq
where the parameter $\nu = 2c^*a^3$ is the nominal volume fraction of
the ions in the electrolyte~\cite{kilic2007a}.

It is convenient to introduce also the average ion or ``salt" concentration and (half) the charge density
\beq[salt:charge]
c = \frac{1}{2}(c_+ + c_-) \quad \mbox{and}\quad \rho = \frac{1}{2}(c_+-c_-),
\eeq
in terms of which the transport equations can be rewritten as
\begin{align}
\partial_t c &= -\epsilon\partial_x F, \label{eq:nernst:planck:salt} \\
\partial_t \rho &= -\epsilon\partial_x J. \label{eq:nernst:planck:charge}
\end{align}
Here $F = \frac{1}{2}(F_+ + F_-)$ and $J = \frac{1}{2}(F_+ - F_-)$ are the average salt flux and current density, respectively,
\begin{align}
%F &= -\partial_x c - \rho\partial_x \phi - \nu c\partial_xc / (1-\nu c), \\
F &= -\partial_x c/(1-\nu c) - \rho\partial_x \phi, \\
J &= -\partial_x \rho - c\partial_x \phi - \nu\rho\partial_xc / (1-\nu c).
\end{align}
Since we assume blocking electrodes we have no-flux boundary condition at the electrodes,
\beq[noflux:bc]
F_\pm = 0 \quad \mbox{or} \quad F = J = 0,
\eeq
whereas the compact layer b.c. reduces to
\beq[stern:bc]
V_\mathrm{ext}-\phi = \mp\epsilon\delta\partial_x\phi \quad \mbox{at} \quad x=\pm 1.
\eeq
Here $V_\mathrm{ext}(t) = \mp V \sin(\omega t)$ is the electrode potential and $\delta =
C_D/C_S$ is the ratio of the compact-layer capacitance $C_S = \varepsilon_S/h_S$ to that
of the diffuse layer $C_D = \varepsilon/\lambda_D$ in the low voltage limit.

%%%%%%%%%%%%%%%%%%%%%%%%%%%%%%%%%%%%%%%%%%%%%%%%%%%%%%%%%%%%%%%%%%%%%%%%%%%%%%%
\subsection{ Dimensionless parameters }

The PNP model contains three dimensionless parameters: $\epsilon$,
$V$, and $\delta$. In aqueous electrolytes, the screening length
has submicron scale, $\lambda_D\approx 1-100$ nm, so the
diffuse-charge boundary layers near the electrodes typically have a
very small dimensionless width $\epsilon = \lambda_D/L \ll 1$, at
least in microsystems where $L \gg 1$~$\mu$m. Although $\epsilon > 1$ is
possible in nanosystems, we restrict our attention to the typical case
$\epsilon \ll 1$, which is the basis for our asymptotic
analysis.

Contrary to most prior work, we focus on the nonlinear regime of large
applied voltages, $V \gg 1$ (or, with units, $V \gg kT/e \approx 25$ mV),
as in Refs.~\cite{bazant2004,bazant2005,chu2005,chu2006,kilic2007a,kilic2007b}. Since
applied voltages larger than a few volts tend to trigger Faradaic
reactions in aqueous electrolytes at ac frequencies around the inverse
$RC$ time~\cite{studer2004,garcia2006}, we envision experimentally relevant values
of $V \approx 1-200$, although larger voltages can be sustained at
higher frequencies or in non-aqueous solvents or liquid salts. Unlike
prior work, we allow for large enough voltages that the double-layers
lose their quasi-equilibrium structure.

The parameter $\delta = \varepsilon h_S/\varepsilon_S\lambda_D = \lambda_S/\lambda_D$ (where $\lambda_S = h_S\varepsilon / \varepsilon_S$ is an effective thickness for the compact layer) can be estimated in some cases, but it is usually adjusted to fit experimental data. For example, let us consider different surfaces in contact with a 1~mM aqueous electrolyte with $\lambda_D = 10 $ nm. For a thin dielectic coating, such as a natural TiO$_2$ oxide layer with $h_S=4$~nm and $\varepsilon_S = 110$ where a constant $C_S$ seems reasonable, we get $\delta\approx0.3$, although much thicker dielectric layers yielding $\delta \approx 10$ can arise in patterned microsystems~\cite{cahill2004}. In nonlinear electrokinetic phenomena, the inferred value of $\delta$, required  to match the standard dilute-solution model to experimental data,  can be up to several orders of magnitude larger, although this is more likely due to failures of the model, and not directly related to surface capacitance~\cite{large2}.

With a dielectric electrode coating, the surface capacitance is relatively clear, but in the classical picture of the Stern model~\cite{stern1924}, it is associated with a hypothetical flat monolayer of water molecules, which limit the approach of hydrated ions at the metal/electrolyte interface. In that case, one would expect $h_S\approx 1$~\AA, and, since alignment of water dipoles is assumed to reduce the permittivity in the Stern layer to $\varepsilon_S\approx 0.1 \varepsilon$ ~\cite{bockris_book,delahay_book}, we estimate $\delta \approx 0.1$. The Stern picture, however, is complicated (at least) by electronic boundary layers in the metal~\cite{kornyshev1982}, chemisorption from the solution~\cite{damaskin1995}, nanoscale surface roughness~\cite{daikhin1996,daikhin1997}, and crowding effects absent in the PNP model of the diffuse layer, all of which can be misattributed to the Stern layer, as emphasized by Bazant et al.~\cite{large2}.  Even for a smooth liquid-mercury electrode, the inferred Stern layer capacitance is voltage dependent~\cite{bockris_book,damaskin1995}. Nevertheless, since our goal here is to analyze the nonlinear dynamics of ions in solution, we will simply assume a constant surface capacitance and allow for a wide range of values $\delta = 0.01 - 10$.

In the simple MPNP model there is one more dimensionless parameter,
$\nu = 2 a^3 c^*$, which controls the importance of crowding
effects. The nominal concentration $c^*$ could range from a very
dilute 1~$\mu$M solution with $6\times10^{20}$ ions/m$^3$ to a
concentrated 1~M solution (near physiological salt levels) with
$6\times10^{26}$ ions/m$^3$.  A natural choice for the effective
molecular lattice spacing $a$ is the diameter of a hydrated ion,
around $4-5$~\AA{} for small ions in water, which would yield $\nu
\approx 10^{-7} - 10^{-1}$.
Taking into account the under-estimation
of steric effects in a hard-sphere liquid by our lattice-based
model~\cite{biesheuvel2007,large2}, the value of $a$ could be increased by
roughly a factor of two~\cite{storey2008}. Electrostatic
correlations also become important when ions are crowded at this
scale, comparable to the Bjerrum length, $7$~\AA{} in bulk water. As a
crude approximation, therefore, we may consider $\nu$ as large as
$0.4$ in a concentrated electrolyte.

\begin{figure}
\includegraphics[width=0.45\textwidth]{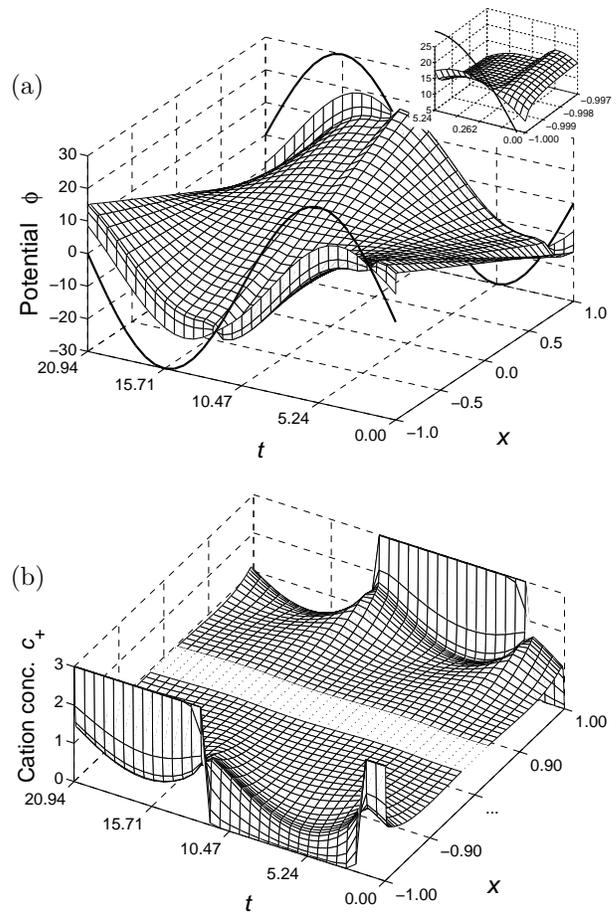}
\caption{Numerical solution of the PNP eqs. for $V = 30$, $\omega = 0.3$, $\delta = 0.3$, and $\epsilon = 0.001$. (a) Potential variation in time and space; solid black line shows the external potential on the electrodes $V_\mathrm{ext}$. The inset zooms onto the rapid potential variation across the diffuse screening layer. (b) Zoom on cation concentration near the electrodes; anion concentration is identical but phase shifted one half period in time.}
\label{fig:pnp30}
\end{figure}

\begin{figure}
\includegraphics[width=0.45\textwidth]{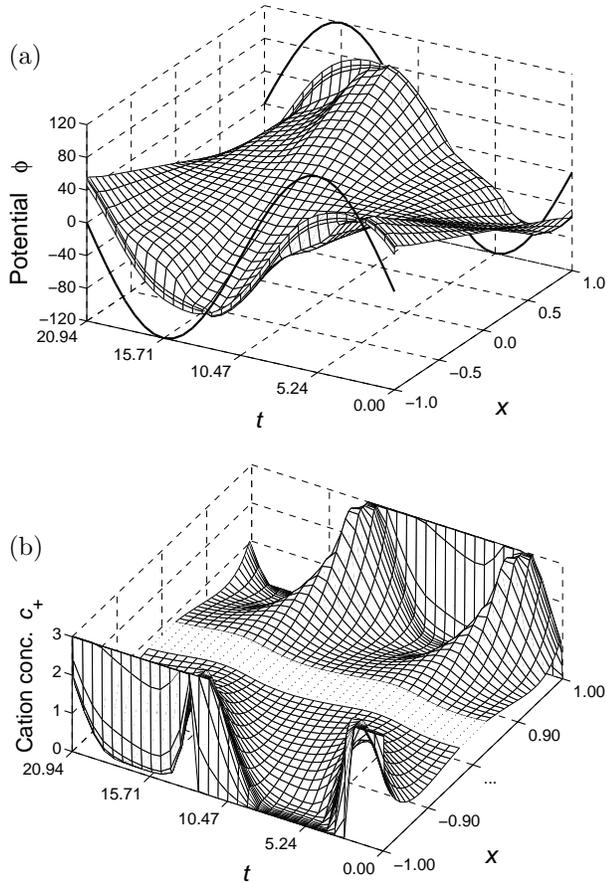}
\caption{Numerical solution of the PNP eqs. for $V = 120$, $\omega = 0.3$, $\delta = 0.3$, and $\epsilon = 0.001$. (a) Potential variation in time and space. (b) Zoom on cation concentration near the electrodes.}
\label{fig:pnp120}
\end{figure}

%%%%%%%%%%%%%%%%%%%%%%%%%%%%%%%%%%%%%%%%%%%%%%%%%%%%%%%%%%%%%%%%%%%%%%%%%%%%%%%
\subsection{Numerical solution\label{sec:numerical:pnp}}

Before embarking on our asymptotic analysis, we present some numerical solutions of the PNP model for the problem sketched in Fig.~\ref{fig:capacitor:sketch}, which will be used to test and calibrate various analytical approximations below. As noted above, steric effects in the MPNP model tend to reduce nonlinearities, so the PNP model serves as a more stringent test case.

We use the \Comsol{} finite element package~\cite{comsolAB} to solve numerically the PNP model in the form of Eqs.~\eqref{eq:poisson}, \eqref{eq:nernst:planck:salt}, and~\eqref{eq:nernst:planck:charge} with boundary conditions \eqref{eq:noflux:bc} and \eqref{eq:stern:bc}. It is necessary to use a very fine mesh of $\Delta x_\mathrm{min} \approx 10^{-5}$ close to the electrodes in order to resolve the highly compressed (and unphysical) diffuse-layer structure in the PNP model, even at $\epsilon = 10^{-3}$. For our 1D problem, this is straightforward to achieve using a nonuniform graded mesh, but in 2D or 3D it would pose a serious problem. In fact, overcoming such limitations is a major motivation for our development of accurate boundary-layer approximations below.
The steady-state periodic response is obtained by integrating forward in time, using the default time-dependent solver of \Comsol{}. Since the transient diffusive relaxation in the bulk is slow, it is necessary to integrate for a very long time, up to 100 times the period of the driving voltage or more, before the steady-state periodic solution is reached.

Figure~\ref{fig:pnp30} shows the result for $V = 30$, $\omega = 0.3$, $\delta = 0.3$, and $\epsilon = 0.001$: Fig.~\ref{fig:pnp30}(a) shows the potential $\phi(x,t)$, and Fig.~\ref{fig:pnp30}(b) displays the cation concentration profile $c_+(x,t)$.
In the bulk region both the cation and anion concentrations are constant and (very close to) unity, and the electrolyte therefore behaves like an ideal resistive medium with unit conductivity. The potential varies linearly throughout the bulk region, driving a constant ohmic current density, and shows a roughly harmonic time variation that is about 45$^\circ$ ahead of the external potential $V_\mathrm{ext} = \pm V\sin(\omega t)$ (solid black line at $x=\mp1$).

In the diffuse screening layer close to the electrodes the ion concentration varies very rapidly from a maximum of $\max\{c_\pm\}\approx3\times10^3$ at the electrode surface down to around unity over a distance of $O(\epsilon)$. The inset in Fig.~\ref{fig:pnp30}(a) zooms onto the rapid potential variation in the screening layer, and the visible difference between the potential on the electrodes and in the electrolyte corresponds to the compact layer voltage, cf.
Eq.~\eqref{eq:stern:bc}.

Within a distance of about 0.1 from the electrodes we see a nonuniform pattern in the cation concentration profile that oscillates at twice the driving frequency, $2\omega$. The concentration has a minimum just about the time when the screening layer is fully charged, and a maximum when the screening layer changes polarity (occurs for $t\approx 1.75$ and again at 12.25). The anion concentration shows a fully similar pattern so that effectively this ``diffusion layer" is charge neutral.

Figure~\ref{fig:pnp120} shows the solution for $V = 120$ with otherwise the same parameters as in Fig.~\ref{fig:pnp30}. The overall picture is essentially the same as before: The bulk ionic concentrations are constant, but due to the massive accumulation of ions around the electrodes (maximal concentration exceeds $4\times10^4$), the bulk concentration is down to 0.86, i.e., 14\% below the nominal value.

An interesting feature is seen in Fig.~\ref{fig:pnp120}(b) for $t \approx 5.24$:
Close to the electrode at $x=-1$ there is an extended region where the cation
concentration drops to zero, while at the same time the anion
concentration is also low but clearly nonzero (see
Fig.~\ref{fig:profiles120} below for a more detailed view). This transient ``space-charge layer"
is similar to the steady counterpart described by
Rubinstein and Shtilman for the case of dc Faradaic conduction when an
electrochemical cell is driven above the diffusion limited current~\cite{rubinstein1979,chu2005}.  It is
also clear from Fig.~\ref{fig:pnp120}(a) that there is a significant
potential drop across the space-charge layer.

%%%%%%%%%%%%%%%%%%%%%%%%%%%%%%%%%%%%%%%%%%%%%%%%%%%%%%%%%%%%%%%%%%%%%%%%%%%%%%%
\section{Asymptotic analysis\label{sec:asymptotics}}

\subsection{Nested boundary layers}

In the limit of thin double layers, $\epsilon =
\lambda_D/L\ll 1$, the dynamical problem can be analysed by matched
asymptotic expansions~\cite{bazant2004}. The standard procedure begins
by seeking regular expansions in the form of power series
\beq
c =
c^{(0)} + \epsilon c^{(1)} + \epsilon^2 c^{(2)} + \ldots,
\eeq
substituting into the governing equations, and collecting like powers
of $\epsilon$. This procedure is guaranteed to converge in the
limit $\epsilon\to 0$ with all other parameters held fixed. However,
for any fixed $\epsilon>0$ there could be $\epsilon$-dependent
restrictions on the other parameters, in particular the driving voltage $V$,
for a truncated expansion to produce accurate
results. Following Bazant \emph{et~al.}~\cite{bazant2004,chu2006} we denote
the regime where such conditions hold as ``weakly nonlinear", as
opposed to the ``strongly nonlinear" regime where the standard
asymptotic expansions breaks down. It is on this ``strongly nonlinear"
regime that we focus our attention. We aim at deriving the
leading-order dominant balance in the joint limit $\epsilon\to 0$ and
$V\to\infty$, and since we focus
exclusively on the leading-order approximation, we drop the superscript
$^{(0)}$ on all variables in order to simplify the notation.

For small applied voltages, it is well known that in the limit
$\epsilon \to 0$ the diffuse part of the double layer acts as a
mathematical boundary layer of $O(1)$ non-zero charge density and
$O(\epsilon)$ thickness on the leading-order quasi-electroneutral bulk region
at the $O(1)$ length scale of the geometry. This is the mathematical
justification for linear circuit models. In the case of blocking
electrodes, the characteristic $RC$ time scale for charging of the
double layers is $O(1)$ in our dimensionless units.

The application of a large voltage leads to the new effect of salt
adsorption by the diffuse layer and related depletion of the bulk
concentration, first described by Bazant \emph{et~al.}~\cite{bazant2004}, and in
higher dimensions, surface conduction through the diffuse layer
becomes important at the same time~\cite{chu2006}. For a suddenly
applied dc voltage at blocking electrodes, during the initial $RC$
charging phase over $O(1)$ time, a thin quasi-electroneutral
diffusion layer extends to $O(\sqrt{\epsilon})$ width.
The next phase of relaxation proceeds at the slow
$O(\epsilon^{-1})$ time scale for bulk diffusion, as concentration gradients
spread across the cell to $O(1)$ distances.

For a large applied ac voltage, this picture is altered by the imposed time scale. In the case of ac forcing close the $RC$ time scale, $\omega^{-1} = O(1)$, the oscillating voltage generally leads to the formation of a thin, nested \emph{oscillating diffusion layer} confined to \emph{steady} $O(\sqrt{\epsilon})$
thickness. Since salt adsorption by the inner diffuse layer is positive,
regardless of the polarity, the diffusion layer oscillates at twice
the driving frequency. It is also accompanied by a gradual bulk salt depletion that propagates across the cell over
$O(1)$ times after the ac voltage is turned on, similar to the case of the sudden dc voltage. However,
in this work, we ignore such initial transients and focus on the
steady ac response after the bulk has relaxed to steady state.

\begin{figure}
\includegraphics[width=0.4\textwidth]{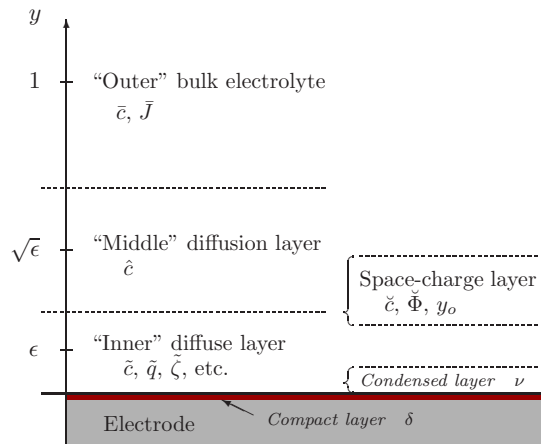}
\caption{(Color online) Schematic picture of nested boundary layers in matched asymptotic expansion:
The bulk ``outer" region is connected via the ``middle" diffusion layer to the ``inner" diffuse layer. A compact (Stern) layer separates the electrolyte from the blocking electrode. At large voltage, local salt depletion in the diffusion layer can cause the double layer to change to a non-equilibrium structure, with an extended ``space-charge" layer that is completely depleted of coions. Further, when the concentration in the diffuse layer approaches the steric limit, a condensed phase of ions forms at the electrode. The figure also indicates some of the variables and parameters introduced in the asymptotic analysis of the different layers.}
\label{fig:nested}
\end{figure}

In this way, we are lead to analyze a new, nested boundary-layer structure sketched in Figure~\ref{fig:nested}, consisting of the ``outer" bulk region (unit length scale), a ``middle" diffusion layer ($\sqrt{\epsilon}$ length scale), and the ``inner" diffuse part of the double layer ($\epsilon$ length scale). This picture remains valid until the voltage becomes large enough to fully deplete the middle diffusion layer, leading to the formation of transient space-charge layers extending by $O(\sqrt{\epsilon})$ or more into the cell, twice per ac period. Our goal in the rest of the paper is to develop uniformly valid asymptotic boundary-layer approximations in all of these cases. For clarity, we denote asymptotic approximations for each region by different accents, as indicated Figure ~\ref{fig:nested}. For example, the salt concentration $c$ is asymptotic to $\bar{c}$ in the bulk, $\hat{c}$ in the diffusion layer, $\tilde{c}$ in the diffuse charge layer, and $\breve{c}$ in the space-charge layer.

%%%%%%%%%%%%%%%%%%%%%%%%%%%%%%%%%%%%%%%%%%%%%%%%%%%%%%%%%%%%%%%%%%%%%%%%%%%%%%%
\subsection{Quasi-electroneutral bulk\label{sec:electroneutral:bulk}}

We begin by analyzing the solution in the bulk region. The Poisson equation~\eqref{eq:poisson} shows that the charge density vanishes to both zeroth and first order in $\epsilon$ so that at leading order
\beq
\bar c_+ = \bar c_- = \bar c,
\eeq
denoting bulk variables by a bar accent. The bulk salt concentration displays diffusive dynamics on the time scale $\bar t = \epsilon t$~\cite{bazant2004}, but on the $RC$ time scale the concentration profile is constant in time $\bar c = \bar c(x)$. Here we focus on the steady state after the bulk transients have relaxed, and by symmetry of our simple model problem, the bulk concentration is then simply constant in space, $\bar c = \bar c_o$. The leading order potential varies linearly in space
\beq
\bar\phi = -\frac{\bar J(t)}{\bar c_o}x,
\eeq
where $\bar J(t)$ is the ohmic current and $\bar c_o$ acts as the bulk conductivity.

In the weakly nonlinear regime the bulk concentration is at the nominal value, $\bar c_o = 1$, whereas in the strongly nonlinear regime the adsorption of ions in the double layers may be so strong as to induce $\bar c_o < 1$.

The problem has the following symmetries about the origin
\beq
\left.
\begin{array}{rcl}
\phi(x,t) &=& -\phi(-x,t), \\
\rho(x,t) &=& -\rho(-x,t), \\
c(x,t) &=& \phantom{+}c(-x,t).
\end{array}
\right\}
\eeq
In Sections \ref{sec:equilibrium:dl}-\ref{sec:diffusion:layer} below we focus on the nested boundary layers developing at the \emph{left} electrode, and for convenience we therefore perform a change of variables, $y=1+x$, such that $y = 0$ corresponds to the electrode surface and $y\geq 0$ to the interior of the cell.

%%%%%%%%%%%%%%%%%%%%%%%%%%%%%%%%%%%%%%%%%%%%%%%%%%%%%%%%%%%%%%%%%%%%%%%%%%%%%%%
\subsection{Quasi-equilibrium double layer\label{sec:equilibrium:dl}}

The singular perturbation in the Poisson equation~\eqref{eq:poisson} gives rise to a boundary layer of width $O(\epsilon)$ where the charge density is nonzero to zeroth order in $\epsilon$. Introducing a scaled spatial variable $\tilde y = y/\epsilon$ to remove the singular perturbation, we can seek regular asymptotic expansions (denoted by tilde accents) in the ``inner" diffuse layer. Substituting into Eqs.~\eqref{eq:mass:conservation} and \eqref{eq:flux:gradient} and using $\partial_{\tilde y} = \epsilon\partial_y$, we find that the double layer is in quasi-equilibrium at leading order with constant electrochemical potential $\tilde\mu_\pm$ across it. The value of $\tilde\mu_\pm$ is determined by matching with the solution in the adjacent quasi-electroneutral diffusion layer
\beq
\tilde\mu_\pm = \lim_{\hat y\to 0}\{\log\hat c \pm \hat\phi\}.
\eeq
The quasi-equilibrium arises because the diffuse charge dynamics relaxes on the Debye time scale $\tilde t = t/\epsilon$, which is much faster than the $RC$ charging time.
The ion distributions are determined from Eq.~\eqref{eq:electrochemical:steric} as
\beq[ion:distribution:steric]
\tilde c_\pm = \frac{\hat c_s\,e^{\mp\tilde\psi}}{1+\nu\hat c_s(\cosh\tilde\psi -1)},
\eeq
where $\tilde\psi = \tilde\phi - \hat\phi$ is the excess potential in the double layer relative to the diffusion layer, and $\hat c_s$ is the limiting value of the salt concentration $\hat c_s = \lim_{y\to0}\hat c$ as seen from the double layer. The excess potential satisfies the MPB equation
\beq
\partial_{\tilde y}^2\tilde\psi = \frac{\hat c_s\sinh\tilde\psi}{1+\nu\hat c_s(\cosh\tilde\psi-1)},
\eeq
which can be integrated once to get the field~\cite{freise1952,kilic2007a}
\beq[dl:field:steric]
\partial_{\tilde y}\tilde\psi = -\mathrm{sign}(\tilde\psi)\sqrt{2\log[1+2\nu\hat c_s\sinh^2(\tilde\psi/2)]/\nu}.
\eeq
Note that the ion concentrations in Eq.~\eqref{eq:ion:distribution:steric} are bounded above by steric exclusion, $\tilde c_\pm \leq 2/\nu$, while at much lower concentrations they reduce to the usual results from dilute theory: In this limit ($\nu \to 0$) the ion profiles are given by the Boltzmann equilibrium distribution,
\beq[boltzmann]
\tilde c_\pm = \hat c_s e^{\mp\tilde\psi},
\eeq
and we obtain the standard PB equation
\beq
\partial_{\tilde y}^2\tilde\psi = \hat c_s\sinh\tilde\psi,
\eeq
yielding the familiar Gouy--Chapman (GC) solution
\beq[gouy:chapman]
\tilde\psi = 4\tanh^{-1}\!\Big[\tanh\!\big(\tilde\zeta/4\big)e^{-\sqrt{\hat c_s}\,\tilde y}\Big].
\eeq
Here the integration constant $\tilde\zeta = \tilde\psi(0)$ is simply the leading order zeta potential, and $1/\sqrt{\hat c_s}$ is the local effective Debye length.

%%%%%%%%%%%%%%%%%%%%%%%%%%%%%%%%%%%%%%%%%%%%%%%%%%%%%%%%%%%%%%%%%%%%%%%%%%%%%%%
\subsection{Surface conservation laws\label{sec:conservation:laws}}

The redistribution of ions across the diffuse layer is instantaneous on the $RC$ time scale, but the total amount of ions absorbed can change only by flux into the layer from the adjacent diffusion layer. Following Bazant et al.~\cite{bazant2004,chu2006,chu2007,kilic2007a,kilic2007b} we quantify this by considering the excess amount of each ionic species accumulated in the double layer, $w_\pm = \epsilon\tilde w_\pm$, where
\beq[surface:integrals]
\tilde w_\pm = \frac{1}{\epsilon}\int_\textrm{d.l.}\big(\tilde c_\pm - \hat c_\pm\big)\,\id y = \int_0^\infty(\tilde c_\pm - \hat c_\pm)\,\id\tilde y.
\eeq
The time evolution of $\tilde w_\pm$ is then determined by
\begin{align}
\partial_t \tilde w_\pm
 &= \int_0^\infty \partial_t\big(\tilde c_\pm - \hat c_\pm\big)\, \id\tilde y = -\lim_{\tilde y\to\infty}\tilde F_\pm \\
 & = -\lim_{\hat y\to0}\hat F_\pm, \label{eq:surface:conservation:law}
\end{align}
where the last equality is obtained by flux matching between the double layer and diffusion layer. We also define the diffuse charge and excess salt concentration by
\beq
\tilde q = \frac{1}{2}(\tilde w_+-\tilde w_-) \quad \mbox{and} \quad \tilde w = \frac{1}{2}(\tilde w_++\tilde w_-).
\eeq
Since the diffusion layer is quasi-electroneutral at leading order, the double-layer charging process is coupled directly to the bulk electric current
\beq[charging]
\partial_t\tilde q = -\bar J(t).
\eeq
Variations in the excess salt $\tilde w$ are coupled to the dynamics in the diffusion layer
\beq[flux:injection]
\partial_t\tilde w = -\lim_{\hat y\to0}\hat F \equiv -\tilde F_o,
\eeq
where the flux injection $\tilde F_o(t)$ at the inner ``edge" of the diffusion layer should be understood as the \emph{driving force} behind the oscillations in the salt concentration in the diffusion layer.  These relations exemplify the general mathematical theory of surface conservation laws, in which the total excess concentrations in a diffuse interface are coupled to normal (and surface) fluxes in a concentrated solution~\cite{chu2007}.

%%%%%%%%%%%%%%%%%%%%%%%%%%%%%%%%%%%%%%%%%%%%%%%%%%%%%%%%%%%%%%%%%%%%%%%%%%%%%%%
\subsection{Oscillating diffusion layer\label{sec:diffusion:layer}}

The diffuse screening layers at the electrodes periodically absorb and expel an excess amount of ions from the surrounding electrolyte. However, the bulk transport of neutral salt is essentially a diffusion process on the time scale $\bar t = \epsilon t$, which is much slower than the ac driving that we consider here, and hence the leading order dynamics are confined to a diffusion layer of $O(\sqrt{\epsilon})$ width around the electrode~\cite{bazant2004,chu2006}. We therefore introduce a scaled spatial variable $\hat y = y/\sqrt{\epsilon}$ and seek regular asymptotic expansions (denoted by hat accents) in this ``middle" diffusion layer.
Substituting into the Poisson equation~\eqref{eq:poisson} we find that, like in the bulk, the charge density vanishes to zeroth order in $\epsilon$, so that the leading order ion concentrations are equal
\beq
\hat c_+ = \hat c_- = \hat c.
\eeq
Equation~\eqref{eq:nernst:planck:salt} then reduces to a simple diffusion problem
\beq
\partial_t\hat c = \partial_{\hat y}^2\hat c,
\eeq
to be solved on the interval $\hat y\in [0,\infty)$. Matching to the bulk solution requires $\lim_{\hat y\to\infty}\hat c \sim \lim_{y\to0}\bar c = \bar c_o$, whereas the boundary condition at the inner edge of the diffusion layer is determined by matching with the salt flux out of the double layer, cf. Eq.~\eqref{eq:flux:injection}
\beq[diffusion:bc]
-\frac{1}{\sqrt{\epsilon}}\lim_{\hat y\to0}\partial_{\hat y}\hat c = \lim_{\hat y\to0}\hat F = \tilde F_o.
\eeq
The solution to the 1D diffusion problem can be expressed in terms of a convolution integral~\cite{bazant2004}
\beq[convolution:infty]
\hat c = \bar c_o + \sqrt{\epsilon}\int_{-\infty}^tG(\hat y,t-t') \tilde F_o(t')\,\id t',
\eeq
where
\beq[transient:kernel]
G(\hat y,t) = \frac{1}{\sqrt{\pi t}}e^{-\hat y^2/4t}
\eeq
is the Green's function for the diffusion equation with a sudden unit flux at $t = 0^+$ injected at the boundary,
\beq
G(\hat y,0) = 0, \quad -\partial_{\hat y}G(0^+,t) = \delta^+(t).
\eeq
for a semi-infinite domain. Technically, Eq.~(\ref{eq:transient:kernel}) is the first term in an expansion for the Green function in a finite bulk domain (Eq. 24 of Ref.~\cite{bazant2004}), which would be needed to describe the initial transient when the ac voltage is first turned on. Here, we focus on the steady-state response, after initial diffusion layers have relaxed across the cell, thereby lowering the uniform bulk concentration $\bar{c}$ (see below), and the oscillating diffusion layers have only $O(\sqrt{\epsilon})$ width, consistent with the semi-infinite approximation (\ref{eq:transient:kernel}).

To describe this situation, since the flux injection is periodic we may rewrite Eq.~\eqref{eq:convolution:infty} as
\beq[convolution:periodic]
\hat c = \bar c_o + \sqrt{\epsilon}\int_0^TG_\omega(\hat y,t-t') \tilde F_o(t')\,\id t' ,
\eeq
where $T = 2\pi/\omega$ is the driving period and
\beq[periodic:kernel]
G_\omega(\hat y,t) = \frac{1}{T}\bigg[-\hat y + \sum_{n=1}^\infty\frac{1}{\sqrt{in\omega}}\,e^{in\omega t-\sqrt{in\omega}\hat y} + \mathrm{c.c.}\bigg]
\eeq
is the Green's function for a periodic influx of salt,
\beq
\langle G_\omega(0,t) \rangle = 0, \ -\partial_{\hat y}G_\omega(0^+,t) = \sum_{n=-\infty}^\infty\delta^+(t-nT).
\eeq
Equations Eqs.~\eqref{eq:convolution:infty} or \eqref{eq:convolution:periodic} clearly show that in the weakly nonlinear regime, where $\tilde F_o$ is $O(1)$, the concentration in the diffusion layer $\hat c$ is equal to the bulk $\bar c$ at leading order; the flux injection only gives rise to an $O(\sqrt{\epsilon})$ perturbation. The strongly nonlinear regime is essentially \emph{defined} as the regime of driving voltages high enough that $\tilde w$ and $\tilde F_o$ grows to $O(1/\sqrt{\epsilon})$ and the variations in $\hat c$ reach $O(1)$.

Since the diffusion layer is charge neutral at leading order, the current is constant across it and equal to the bulk current $\bar J(t)$. However, the conductivity differs from its bulk value, which gives rise to {\it transient concentration polarization}. There is an excess electrostatic potential variation $\hat\psi = \hat\phi - \bar\phi$, and an excess field given by
\beq[excess:ohmic:field]
-\frac{1}{\sqrt{\epsilon}}\partial_{\hat y}\hat\psi = \bar J\bigg(\frac{1}{\hat c}-\frac{1}{\bar c}\bigg).
\eeq
In the weakly nonlinear regime when $(1/\hat c-1/\bar c)$ is $O(\sqrt{\epsilon})$ it is clear that $\hat\psi$ is only an $O(\epsilon)$ perturbation; in the strongly nonlinear regime $\hat\psi$ grows to $O(\sqrt{\epsilon}\bar J)$ perturbation which is, however, still negligible compared to the bulk $\bar\phi = -\bar J x/\bar c_o$.

Finally, the leading order charge density in the diffusion layer can be evaluated by substituting Eq.~\eqref{eq:excess:ohmic:field} into the Poisson equation to get
\beq[diffusion:charge]
\hat\rho = -\epsilon\partial_{\hat y}^2\hat\psi = -\epsilon^{3/2}\frac{\bar J\partial_{\hat y}\hat c}{\hat c^2}.
\eeq
The quasi-electroneutral solution in the diffusion layer remains valid for $|\hat\rho|\ll \hat c$; we return to this aspect in Sec.~\ref{sec:asymptotics:beyond}.

%%%%%%%%%%%%%%%%%%%%%%%%%%%%%%%%%%%%%%%%%%%%%%%%%%%%%%%%%%%%%%%%%%%%%%%%%%%%%%%
\subsection{Closing the problem\label{sec:dynamical:model}}

In order to close the coupled problem for the dynamical variables $\bar J(t)$, $\tilde q(t)$, $\tilde\zeta(t)$, $\tilde w(t)$, and $\hat c_s(t)$ we need a few more relations between them. The first is obtained by writing the overall potential drop over the boundary layers, from the electrode to the bulk electrolyte at $x=-1$, as the sum of the contributions from the compact and diffuse layers, to get
\beq[potential:drop]
 V_\ext - \bar\phi(-1,t) = V_\ext - \bar J/\bar c_o = - \tilde q\,\delta + \tilde\zeta.
\eeq
Since we focus on the leading order approximation, we neglect here the small potential drop over the diffusion layer. Next, the diffuse-layer voltage $\tilde\zeta$ can be related to the diffuse charge through Eq.~\eqref{eq:dl:field:steric} for the field at the electrode surface, yielding
\beq[excess:charge:steric]
\tilde q = -\mathrm{sign}(\tilde\zeta)\sqrt{2\log[1+2\nu\hat c_s\sinh^2(\tilde\zeta/2)]/\nu},
\eeq
and in the dilute limit this reduces to Chapman's formula
\beq[excess:charge]
\tilde q = -2\sqrt{\hat c_s}\sinh(\tilde\zeta/2).
\eeq
The charge-voltage relation can be inverted to get
\beq[excess:voltage:steric]
\tilde\zeta = -\mathrm{sign}(\tilde q)\,2\sinh^{-1}\sqrt{\frac{e^{\nu\tilde q^2/2}-1}{2\nu\hat c_s}}.
\eeq
The excess salt concentration can be expressed in integral form~\cite{kilic2007b}
\beq[excess:salt:integral]
\tilde w = \int_0^{\tilde\zeta}\frac{\tilde c}{\partial_{\tilde y}\tilde\psi}\,\id\tilde\psi,
\eeq
and using PB theory in the dilute limit the integral can be evaluated to get~\cite{bazant2004}
\beq
\tilde w =  4\sqrt{\hat c_s}\sinh^2(\tilde\zeta/4),
\eeq
or, eliminating $\tilde\zeta$ we obtain
\beq[excess:salt]
\tilde w = \sqrt{\tilde q^2 + 4\hat c_s} - \sqrt{4\hat c_s}.
\eeq
For the MPB model, Eq.~\eqref{eq:excess:salt:integral} is difficult to handle analytically, but numerical integration shows that Eq.~\eqref{eq:excess:salt} approximates the integral well, with relative error of $O(\nu)$.

The bulk salt concentration $\bar c_o$ is determined by imposing the global conservation of salt in the cell
\beq
\int_{-1}^1 c(x,t) \,\id x = 2.
\eeq
As noted in Ref.~\cite{bazant2005}, integral constraints on the total number of inactive ions are generally required for steady-state problems to replace information about the initial condition (e.g. when the voltage is first turned on) that is preserved during time evolution with no-flux boundary conditions.

There is a periodic exchange of salt between the inner diffuse and middle diffusion layers, but Eq.~\eqref{eq:convolution:periodic} shows that $\langle\hat c\rangle = \bar c$, i.e., the diffusion layer does not contain any \emph{excess} salt on time average. As a result, the uniform bulk concentration in the (time-periodic) steady state, $\bar c_o$ , is reduced only by the time-averaged salt adsorption of the diffuse layers,
\beq[cbulk]
\bar c_o = 1 - \epsilon\langle\tilde w\rangle.
\eeq
which also describes the (static) steady state after a sudden dc voltage is imposed~\cite{bazant2004}.
This, together with Eqs.~\eqref{eq:charging}, \eqref{eq:flux:injection}, \eqref{eq:convolution:periodic}, \eqref{eq:potential:drop}, \eqref{eq:excess:voltage:steric}, and \eqref{eq:excess:salt} constitute a set of ``ordinary" integro-differential-algebraic equations in time for the dynamical variables $\bar J(t)$, $\tilde q(t)$, $\tilde F_o(t)$, $\tilde w(t)$, $\hat c_s(t)$, and $\tilde\zeta(t)$.

Our focus in the present work is on the steady-state periodic response, and we explicitly made use of this in deriving our dynamical model by replacing the transient Eq.~\eqref{eq:transient:kernel} with Eq.~\eqref{eq:periodic:kernel}. The problem could be solved numerically by a relaxation method, representing each dynamical variable by a truncated Fourier series, as done in Ref.~\cite{olesen_thesis}. Here, however, our approach is to integrate the dynamical equations by a timestepping algorithm; the integration is continued until a periodic state is reached, typically within 10-20 periods of the driving voltage. The timestepping approach is well suited for integrating also the non-equilibrium model developed in Sec.~\ref{sec:asymptotics:beyond}, and is much more efficient on computer memory for solving problems with 2D or 3D electrode geometry. Further details are given in our supplementary material~\cite{epaps}.

%%%%%%%%%%%%%%%%%%%%%%%%%%%%%%%%%%%%%%%%%%%%%%%%%%%%%%%%%%%%%%%%%%%%%%%%%%%%%%%
\section{Weakly nonlinear regime\label{sec:weakly:nonlinear}}

The ``weakly nonlinear" regime defined in Ref.~\cite{bazant2004} is characterized by fluctuations in the diffusion layer salt concentration being only a small perturbation to the bulk value, so that $\hat c = \bar c = 1$ at leading order. This is the response predicted by matched asymptotic expansions in the singular limit $\epsilon\to0$ with all other parameters held fixed, including $V$. As such the dimensionless, leading-order response is \emph{independent} of $\epsilon$.  We begin with an analysis of this regime, and the findings here form the basis for understanding the peculiarities of the ``strongly nonlinear" regime in subsequent sections, where the solution has a nontrivial dependence on $V$ and $\epsilon$.

%%%%%%%%%%%%%%%%%%%%%%%%%%%%%%%%%%%%%%%%%%%%%%%%%%%%%%%%%%%%%%%%%%%%%%%%%%%%%%%
\subsection{Charge-voltage relation\label{sec:charge:voltage}}

\begin{figure}
\includegraphics[width=0.45\textwidth]{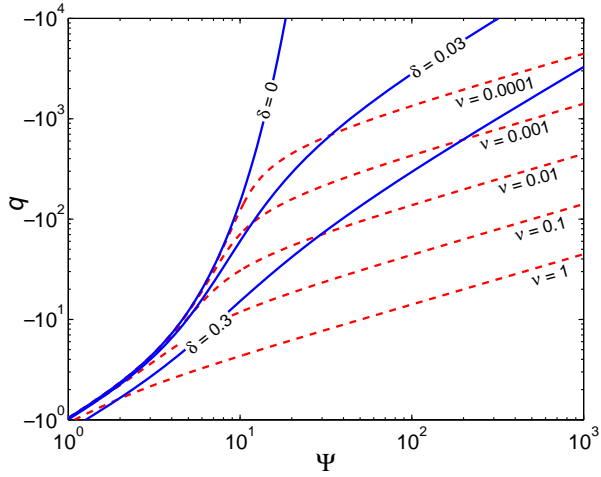}
\caption{(Color online) Quasi-steady (dimensionless) accumulated charge $\tilde q$ in the double layer as a function of its voltage drop $\tilde\Psi = \tilde\zeta-\tilde q\,\delta$, plotted for different values of the Stern parameter $\delta$ with $\nu = 0$ (solid), and different values of the steric parameter $\nu$ with $\delta = 0$ (dashed), all in the weakly nonlinear regime with $\hat c = 1$.}
\label{fig:steric}
\end{figure}

The only nonlinearity in the weakly nonlinear model arises from the diffuse-layer charge-voltage relation, Eq.~\eqref{eq:excess:charge:steric}. In Fig.~\ref{fig:steric} we plot the accumulated charge $\tilde q$ as a function of the overall potential drop $\tilde\Psi = \tilde\zeta - \tilde q\,\delta$ across the double layer for different values of the capacitance ratio $\delta$ and nominal ion volume fraction $\nu$.

In the Debye--H\"uckel limit, $\tilde\zeta \ll1$, Eq.~\eqref{eq:excess:charge:steric} can be linearized to get simply $\tilde q = -\tilde\zeta = -\tilde\Psi/(1+\delta)$. At larger voltage the classical PB theory predicts a dramatic increase in the diffuse-layer capacitance, and $\tilde q$ grows exponentially with $\tilde\zeta$. In Fig.~\ref{fig:steric} this behaviour shows directly on the curve $\delta=\nu=0$ (Gouy--Chapman (GC) model) that bends up sharply for $\tilde\Psi\gtrsim 10$; at $\tilde\Psi = 20$ the concentration in the diffuse layer, cf. Eq.~\eqref{eq:boltzmann}, exceeds $10^8$ times the bulk concentration, which is absurdly high for aqueous electrolytes.

This well-known unphysical artifact of PB theory is alleviated (but not eliminated) in the Gouy--Chapman--Stern (GCS) model by assuming a finite compact layer capacitance, corresponding to a positive value of $\delta$. Then at large voltage the major part is carried by the compact layer, $\tilde\Psi\approx -\tilde q\,\delta$, while the diffuse-layer voltage remains small, $|\tilde\zeta| \approx 2\log|\tilde q| \approx 2\log|\tilde\Psi/\delta|$. This regularizes the problem at moderate voltages, but the success may be misleading: It is unlikely that an \aa ngstr\"om-thick molecular Stern layer could withstand several volts without dielectric breakdown. Moreover, the GCS model does not impose a maximum charge density and at sufficiently large voltages still reaches unphysical ion concentrations.

A more realistic approach could account for crowding effects at large voltages using MPB theory~\cite{kilic2007a,large2}, e.g. leading to Bikerman's model described above~\cite{bikerman1942}. This is equivalent to PB theory at concentrations well below the limit of steric exclusion, $\tilde c_\pm \ll 2/\nu$, as seen clearly in Fig.~\ref{fig:steric}. However, once steric effects saturate the charge density, the diffuse-layer capacitance quickly drops due to the condensed phase of ions forming at the electrode~\cite{stern1924,freise1952,kilic2007a,large2}. This occurs for $\tilde q^2\gtrsim 2/\nu$, and at still larger voltage the overall potential drop is primarily on the condensed layer, with Eq.~\eqref{eq:excess:charge:steric} reducing to
\beq[Csqrt]
|\tilde q| \approx \sqrt{\frac{2|\tilde\zeta|}{\nu}} \approx \sqrt{\frac{2|\tilde\Psi|}{\nu}} \ \mbox{ for } \ |\tilde\Psi| \gg \frac{2}{\nu}
\eeq
Comparing the GCS model for a Stern monolayer with $\delta = 0.03$ to Bikerman's model, the latter predicts (much) lower charging already for $\tilde\Psi\gtrsim 30$ (or 750 mV) at $\nu=10^{-4}$. Even for our example of an oxide layer on the electrodes with $\delta = 0.3$, crowding effects in the liquid could significantly affect the charge-voltage response for $\tilde\Psi \gtrsim 100$ (or 2.5 V), which is still within the range of many experiments.

This asymptotic square-root dependence of the charge-voltage relation (\ref{eq:Csqrt}) is a generic consequence of volume constraints~\cite{large2}, not only in Bikerman's lattice-gas model, but also hard-sphere liquid models, since it corresponds to a diffuse layer of uniform charge density. Once the condensed layer forms, its voltage  $\tilde\zeta\approx \nu\tilde q^2/2$ can easily exceed that of the outer (PB) part of the diffuse layer  $\tilde\zeta \approx 2\log(\tilde q)$, even while the latter remains thicker. In this regime of the model, the thickness of the condensed layer is $\tilde\ell = \ell/\epsilon \approx \nu \tilde q$, which does not become larger than the diffuse-layer thickness until $\tilde q > 1/\nu$.

%%%%%%%%%%%%%%%%%%%%%%%%%%%%%%%%%%%%%%%%%%%%%%%%%%%%%%%%%%%%%%%%%%%%%%%%%%%%%%%
\subsection{Dynamical response\label{sec:time:series:weakly}}

The leading order dynamic response in the weakly nonlinear regime is governed by
\begin{align}
\partial_t\tilde q &= -\bar J, \\
V_\mathrm{ext} - \bar J &= \tilde\zeta - \tilde q\,\delta, \\
%\tilde q &= -\mathrm{sign}(\tilde\zeta)\sqrt{2\log[1+2\nu\sinh^2(\tilde\zeta/2)]/\nu}, \label{eq:excess:charge:weakly} \\
\tilde\zeta &= 2\sinh^{-1}\sqrt{\frac{e^{\nu\tilde q^2/2}-1}{2\nu}} \label{eq:excess:charge:weakly}  \\
\hat c &= \bar c = 1.
\end{align}
This may be rewritten as a single ordinary differential equation for the double-layer voltage $\tilde\Psi$~\cite{bazant2004}
\beq
C\, \partial_t\tilde\Psi = \bar J = V_\mathrm{ext} - \tilde\Psi,
\eeq
where $C(\tilde\Psi) = - \id\tilde q/\id\tilde\Psi$ is the total differential capacitance of the double layer, and $V_\mathrm{ext}(t) = V\sin(\omega t)$ is the external driving voltage.

\begin{figure*}
\includegraphics[width=0.9\textwidth]{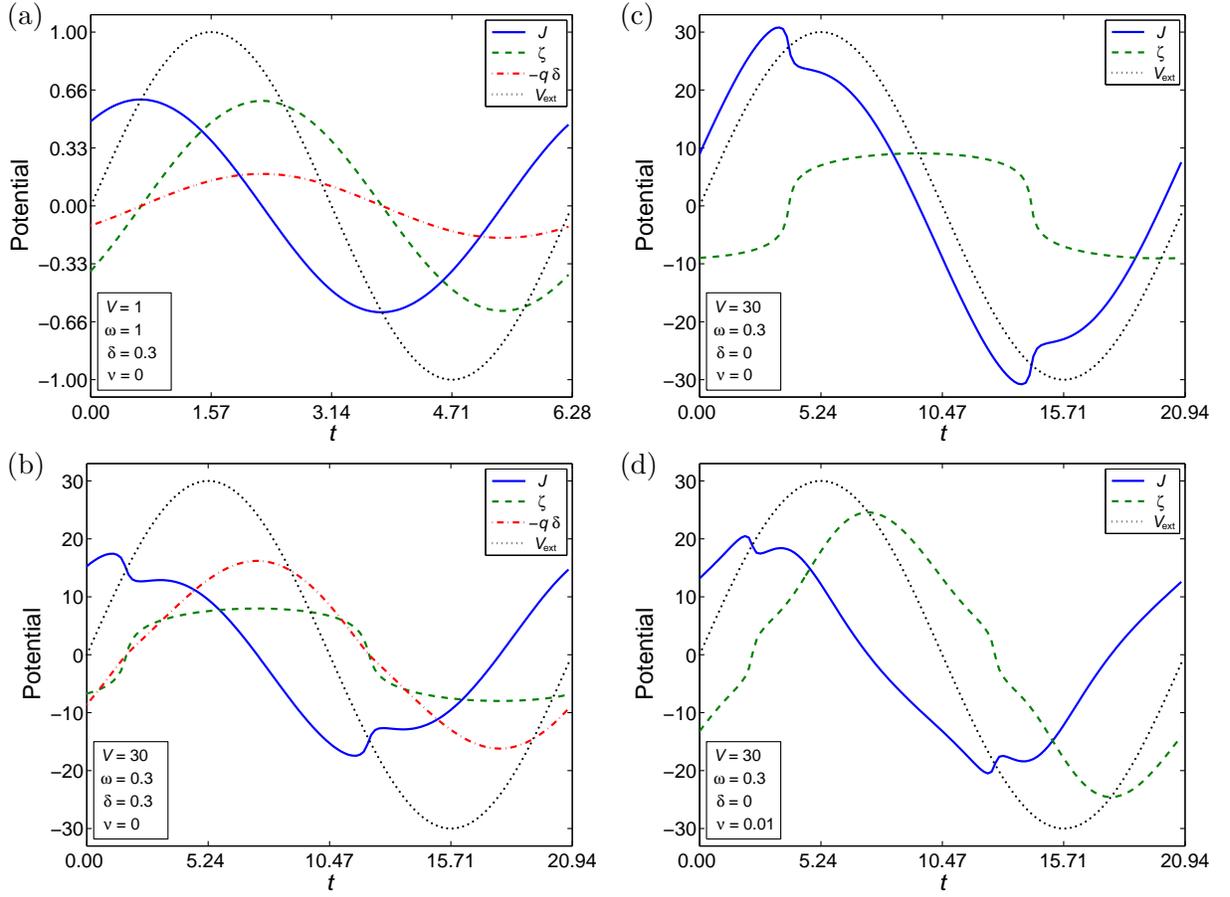}
\caption{(Color online) Distribution of dimensionless cell voltage, $V_\mathrm{ext} = \bar{J} + \tilde\zeta - \tilde q\,\delta$ (dotted), divided into contributions across the bulk electrolyte, $\bar J$ (solid), diffuse layer, $\zeta$ (dashed), and compact layer, $-q\,\delta$ (dash-dot), all scaled to the thermal voltage $kT/ze$, as a function of time for different values of the parameters $V$, $\omega$, $\delta$, and $\nu$. The panels show (a) Debye--H\"uckel limit, (b) Gouy--Chapman--Stern (GCS) model, (c) Gouy--Chapman (GC) model, and (d) Bikerman model.}
\label{fig:signal:weakly}
\end{figure*}

We focus on the periodic response obtained by starting from an initially uncharged state and integrating forward in time until all transients have died out. Figure~\ref{fig:signal:weakly} shows the results for different values of the model parameters:

Fig.~\ref{fig:signal:weakly}(a) shows the result for $V = 1$, $\omega = 1$, $\delta = 0.3$, and $\nu = 0$. At this low voltage the charge-voltage relation is still essentially linear, so the system behaves like a linear $RC$ circuit with time constant $(1+\delta)^{-1}$. The double-layer voltage $\tilde\Psi$ is dominated by the diffuse layer with the compact layer contributing only a small fraction $\delta$.

Figure~\ref{fig:signal:weakly}(b) shows the solution at larger voltage $V = 30$ with $\omega = 0.5$, $\delta = 0.3$, and $\nu=0$. At this voltage the relation between $\tilde\zeta$ and $\tilde q$ is clearly nonlinear, $\tilde\zeta$ stalls for $|\tilde q|\gtrsim 10$, and the double-layer voltage becomes dominated by the compact layer. When the double layer changes polarity this in turns makes the change of sign of $\tilde\zeta$ look like a ``sharp" transition which gives rise to a jump in the bulk current.
Those features are even more pronounced in Fig.~\ref{fig:signal:weakly}(c), showing the corresponding solution for $\delta = 0$, i.e., without any compact layer on the electrodes. The double-layer voltage remains low so the bulk current is almost in phase with the driving voltage.

As discussed in the previous section, the very large capacitance of the diffuse layer predicted by PB theory is not realistic. For the solution in Fig.~\ref{fig:signal:weakly}(c) the maximal ion concentration in the diffuse layer almost reaches $10^4$ times the bulk concentration, which could easily trigger steric effects, even for a nominally dilute electrolyte.
Figure~\ref{fig:signal:weakly}(d) shows the result when such are taken into account with a bulk volume fraction $\nu = 0.01$. The result is markedly different: When steric effects set in, the diffuse-layer capacitance drops and $\tilde\zeta$ grows rapidly with $\tilde q$. At lower charging, though, the system is still governed by dilute theory, so we still see a rapid shift in $\tilde\zeta$ with an associated jump in $\bar J$ when the double layer changes polarity.

%%%%%%%%%%%%%%%%%%%%%%%%%%%%%%%%%%%%%%%%%%%%%%%%%%%%%%%%%%%%%%%%%%%%%%%%%%%%%%%
\subsection{Equivalent circuit\label{sec:impedance}}

\begin{figure}
\includegraphics[width=0.45\textwidth]{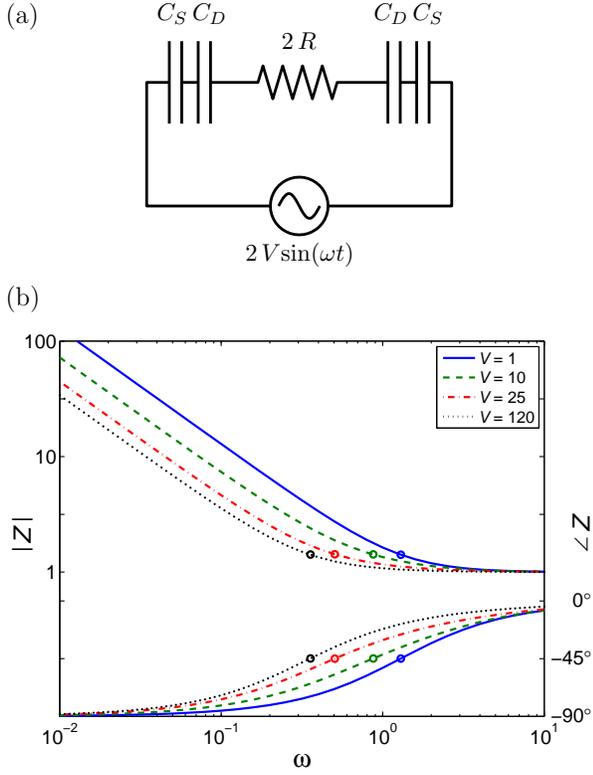}
\caption{(Color online) (a) Equivalent circuit representation for weakly nonlinear dynamics: Compact and diffuse-layer capacitors in series with a bulk resistance. (b) Bode plot of the magnitude $|Z|$ and phase angle $\angle Z$ of the half-cell impedance for increasing driving voltage at $\delta = 0.3$ and $\nu = 0$. The characteristic frequency $\omega_o$, where $\angle Z$ passes through $-45^\circ$ and $|Z|$ bends up, is marked with circles.}
\label{fig:circuit}
\end{figure}

A useful concept for analyzing the cell response is an equivalent circuit diagram like that shown in Fig.~\ref{fig:circuit}(a). The transport through the bulk electrolyte is represented by an ohmic resistor $2R = 2$, and the charge accumulation in the double layer by a series coupling of two capacitors $C_S$ and $C_D$
\beq
\frac{1}{C} = \frac{1}{C_S} + \frac{1}{C_D}.
\eeq
Here $C_S = 1/\delta$ is the capacitance of the compact (Stern) layer, and $C_D = -d\tilde q/d\tilde\zeta$ is the differential capacitance of the diffuse (Debye) layer, given by~\cite{freise1952,kilic2007a,kornyshev2007,large2}
\beq
%C_D = \frac{|\sinh\tilde\zeta|}{[1+\nu(\cosh\tilde\zeta-1)]\sqrt{\frac{2}{\nu}\log[1+2\nu\sinh^2(\tilde\zeta/2)]}}.
%C_D = \frac{|\sinh\tilde\zeta|}{[1+\nu(\cosh\tilde\zeta-1)]\sqrt{2\log[1+2\nu\sinh^2(\tilde\zeta/2)]/\nu}}.
C_D = \frac{|\sinh\tilde\zeta|}{\big[1+2\nu\sinh^2(\tilde\zeta/2)\big] \sqrt{2\log\big[1+2\nu\sinh^2(\tilde\zeta/2)\big]/\nu}}.
\eeq
In the Debye--H\"uckel limit this reduces to $C_D = 1$ and $C = 1/(1+\delta)$. At higher voltage, PB theory predicts a dramatic increase of $C_D = \cosh(\tilde\zeta/2)$, to the extent that $C \approx 1/\delta$. According to MPB theory, the diffuse-layer capacitance becomes a \emph{non-monotonic} function of $\tilde\zeta$, where the initial increase is followed by a decrease as $C_D \approx 1/\nu\tilde q \approx 1/ \sqrt{2\nu\tilde\zeta}$ once steric exclusion sets in.

The equivalent circuit representation is useful for understanding and interpreting the system response. However, from an experimental point of view the overall cell impedance is a key property that can easily be measured with high accuracy, e.g., using a lock-in amplifier. We define the (half) cell impedance $Z$ as the ratio between the first Fourier components of the applied voltage and the resulting current
\beq
Z = \frac{\int_0^T V_\mathrm{ext}(t)\,e^{-i\omega t}\,\id t}{\int_0^T \bar J(t)\,e^{-i\omega t}\,\id t}.
\eeq
Since the system is nonlinear, the impedance so defined is a function of both driving frequency and voltage. Figure~\ref{fig:circuit}(b) shows a Bode plot of the cell impedance $Z$ for different values of $V$ at $\delta = 0.3$ and $\nu = 0$. The curve shape is characteristic of an $RC$ series coupling. At high frequency the ohmic resistance of the bulk electrolyte dominates and $|Z|$ levels off at unity, while at low frequency the double-layer capacitance dominates and $|Z| \propto \omega^{-1}$. At the same time the phase angle $\angle Z$ drops from zero at high frequency to $-90^\circ$ at low frequency. We define the \emph{characteristic frequency} $\omega_o$ for a given driving voltage as that frequency where the phase angle passes through $-45^\circ$, i.e.,
\beq
\angle Z(\omega_o) = -45^\circ.
\eeq
At this frequency the resistive and capacitive components contribute equally much to the overall cell impedance. Figure~\ref{fig:circuit}(b) clearly shows that as the voltage is increased, the double-layer capacitance grows, and the characteristic frequency shifts down.

\begin{figure}
\includegraphics[width=0.45\textwidth]{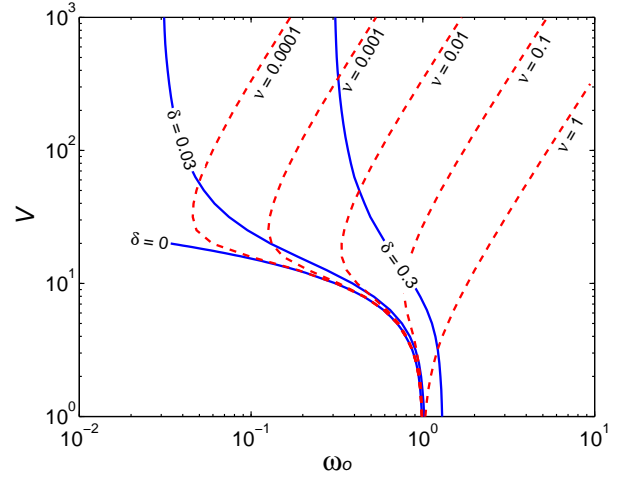}
\caption{(Color online) Characteristic frequency $\omega_o$ vs. driving voltage, plotted for different values of $\delta$ with $\nu = 0$ (solid), and different values of $\nu$ with $\delta = 0$ (dashed).}
\label{fig:rc:time}
\end{figure}

The voltage dependence of $\omega_o$ is shown in more detail in Fig.~\ref{fig:rc:time}, where $\omega_o$ is plotted versus $V$ for different values of $\delta$ and $\nu$. The GCS model simply predicts $\omega_o$ should drop from $\omega_o = 1+\delta$ at low voltage to $\omega_o \approx \delta$ at higher voltage. The same trend is seen for the Bikerman model, up to the point where steric exclusion sets in; beyond this the double-layer capacitance decreases and $\omega_o$ increases, scaling as $\omega_o = O(\sqrt{\nu V})$ at large voltage. These qualitative features predicted by our analysis may be interesting to compare to experimental impedance measurements at large ac voltages, below the threshold for Faradaic reactions or specific adsorption of ions, to seek evidence of steric effects in the liquid phase.
%The scaling of $\omega_o$ with $V$ in this limit is obtained by equating $1/\omega C_D\approx \nu\tilde q/\omega$ with unity and using $\tilde q\sim \bar J/\omega \sim V/\omega$ for $\omega \gtrsim \omega_o$ so that $\omega_o \sim \sqrt{\nu V}$ at large voltage in the Bikerman model.

%The characteristic frequency is important for many applications. For example, ac electroosmosis is maximized around this frequency.

%%%%%%%%%%%%%%%%%%%%%%%%%%%%%%%%%%%%%%%%%%%%%%%%%%%%%%%%%%%%%%%%%%%%%%%%%%%%%%%
\subsection{Neutral salt adsorption }

\begin{figure}
\includegraphics[width=0.45\textwidth]{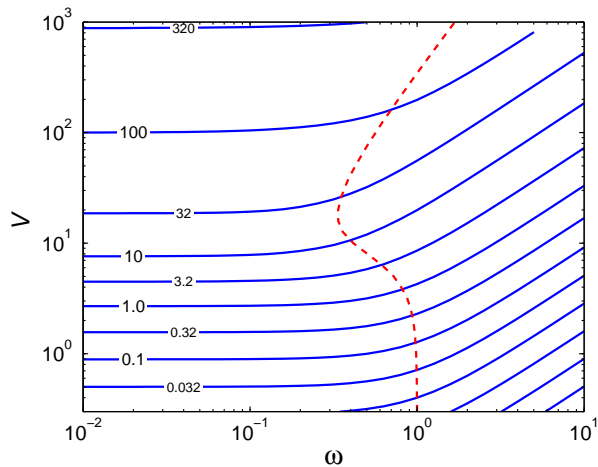}
\caption{(Color online) Contour plot of $\langle\tilde w\rangle$, the time average excess salt concentration in the diffuse layer, as a function of driving frequency and voltage for $\delta=0$ and $\nu=0.01$. The dashed line marks the characteristic frequency $\omega_o$.}
\label{fig:wavg}
\end{figure}

In response to the ac driving, the diffuse layer periodically absorbs and expels an excess amount of ions. At low voltage the charging comes about from both uptake of counterions and expulsion of coions, so the net salt adsorption is low, $\tilde w \approx \tilde q^2/4$, cf. Eq.~\eqref{eq:excess:salt} for $\tilde q\ll1$. At higher voltage there are essentially no more coions to expel, so the charging process is dominated by uptake of counterions and $\tilde w \approx |\tilde q|$.

The excess amount of (counter) ions is taken up from both the adjacent diffusion layer \emph{and} from that at the opposite electrode. In order to estimate when this effect starts to significantly perturb the concentration in the diffusion layer, it is necessary to know the time-average salt uptake $\langle\tilde w\rangle$. This is shown in Fig.~\ref{fig:wavg} as a function of driving voltage and frequency for the Bikerman model with $\delta = 0$ and $\nu = 0.01$. At low frequency, $\omega\ll\omega_o$, the double layer is almost fully charged, so that $\tilde\Psi\approx V_\mathrm{ext}$. At low voltage, $V\lesssim 1$, the figure shows that $\langle\tilde w\rangle \approx V^2/8$, while at high voltage, $V\gtrsim 30$, the steric effects dominate and $\langle\tilde w\rangle \approx \sqrt{V/\nu}$. For comparison, the GCS model predicts $\langle\tilde w\rangle \approx 2V/\pi\delta$ in this limit, and the GC model $\langle\tilde w\rangle \approx \exp(V/2)$. At high frequency, $\omega\gg\omega_o$, the bulk resistance dominates the cell impedance, so $\bar J\approx V_\mathrm{ext}$ and $\tilde q = O(\bar J/\omega)$, from which the scaling is $\langle\tilde w\rangle \approx V^2/8\omega^2 \lesssim 1$ or $\langle\tilde w\rangle \approx 2V/\pi\omega \gtrsim 10$, depending on the level of charging.

%This leads to a net depletion of the leading-order bulk salt concentration, $\bar{c}_0 = 1$ $\epsilon \tilde{w}$, as noted above.

\subsection{The limit of ionic liquids}

The weakly nonlinear regime in a blocking electrolytic cell generally breaks down at large voltages, when the neutral salt adsorption is by the diffuse layers is strong enough to significantly deplete the quasi-neutral bulk solution in the diffusion layers~\cite{bazant2004,chu2006,kilic2007a,kilic2007b}. This phenomenon, however, relies on the availability of available space in the liquid (free of ions) for the total density of ions to become much more concentrated in one region (the double layers) at the expense of another region (the bulk diffusion layers), which is controlled in our MPNP model by the parameter $\nu = 2a^3c^* = 2c^*/c_\mathrm{max}$.  In liquid electrolytes,  $\nu$ represents the bulk volume fraction of (all) solvated ions, which is typically much less than one, and even in saturated solutions of highly soluble ions would rarely exceed 0.1. As such, strongly nonlinear effects must generally be considered (below) in electrolytes at large applied voltages, especially in small systems.

The situation is different in ionic liquids or molten salts, which may be described by the limit $\nu\to 1$ in our MPNP model. This corresponds to the mean-field theory proposed by Kornyshev~\cite{kornyshev2007} where a value $\nu < 1$ could model a somewhat lower volume fraction  of the quasi-neutral  bulk liquid phase, compared to the charged  double layers, where strong normal electric fields may compress the counterions against a charged surface (as described above).  In a molten salt, this density variation may be comparable to the expansion upon melting of an ionic crystal, which can be as large as 20\%, so we might expect $\nu$ to be as small as 0.8, which is still much larger than for a typical electrolyte. This simple approach has had some success in describing experiments and simulations of simple ionic liquids~\cite{federov2008,federov2008b}, in what we would call a weakly nonlinear approximation, where the voltage-dependent quasi-equilibrium double-layer capacitance is coupled to a constant bulk resistor.

\begin{figure}
\includegraphics[width=0.4\textwidth]{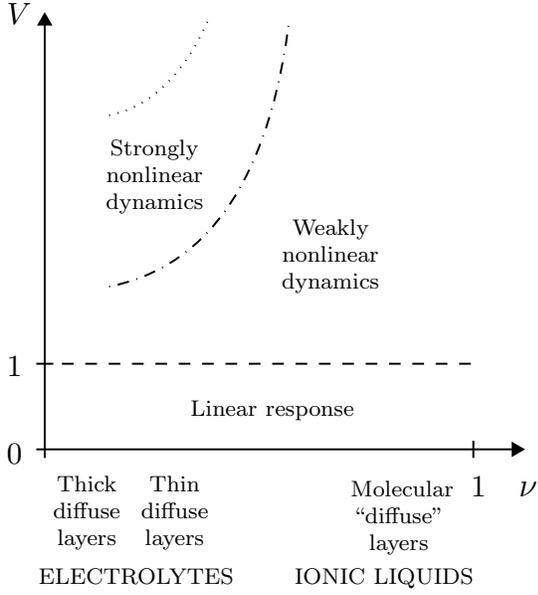}
\caption{Sketch of the different dynamical regimes for a blocking cell in the space of applied voltage, $V$ (scaled to $kT/e$), and nominal bulk volume fraction of ions, $\nu = c^*/c_\mathrm{max}$. Linear response (below the dashed line) holds for $V \ll 1$ for any $\nu$ and diffuse-layer thickness $\epsilon = \lambda_D/L$. For thin diffuse layers ($\epsilon \ll  1$) in electrolytes ($\nu \ll 1$) there is a transition for $V > 1$ to weakly nonlinear dynamics, where the diffuse layer acts as a voltage-dependent capacitor in series with a constant bulk resistance; at larger voltages, there is a transition to strongly nonlinear dynamics, which occurs first only with the oscillating diffusion layers (dash-dot line); at higher voltages there is another transition (dotted line) where the bulk solution becomes uniformly depleted by time-averaged mass transfer into the diffuse layers. The transition curves rise steeply with $\nu$. For ionic liquids and the molten salt limit, $\nu \approx 1$, only the weakly nonlinear regime is possible, since there is not enough volume available to compress significant numbers of ions in the diffuse layers, which approach the molecular scale $a < \lambda_D$.}
\label{fig:ionic}
\end{figure}

An important prediction of our analysis is that this picture always remains valid up to large applied voltages for sufficiently large $\nu$, so that ionic liquids can generally be described by the simple, weakly nonlinear approximation. For a highly concentrated electrolyte (still with $\nu \ll 1$), we can use the estimate of Kilic et al.~\cite{kilic2007b} for the critical voltage $V_c$ (defined by $\epsilon \langle \tilde w \rangle = 1$) to significantly deplete the steady-state bulk salt concentration,
\beq
V_c \approx \frac{\nu}{2\epsilon^2} = \frac{2 (ze)^2 L^2 a^3 c^{*2} }{\varepsilon kT} \propto \left( \frac{c^*}{c_\mathrm{max}}\right)^2
\eeq
which grows with concentration like $\nu^2$ (for fixed ion size). The basic picture is sketched in Fig.~\ref{fig:ionic}, where this transition is represented by the dotted line. Before this transition is reached, the weakly nonlinear approximation breaks down due to significant concentration variations, which are not large enough to deplete the bulk and remain confined to the diffusion layers. We estimate and discusse these transitions below in terms of voltage, but here we note that they also rise steeply with concentration in the limit of ionic liquids, $\nu \to 1$. In a molten salt $\nu \approx 1$, the strongly nonlinear regime disappears, and the nonlinear RC circuit approximation holds for all voltages.

The weakly nonlinear dynamics of ionic liquids in the MPNP model are not very different from those of concentrated electrolytes at large enough voltages to trigger steric effects in the double layer.  In Figs.~\ref{fig:steric} and~\ref{fig:rc:time} we have also included curves for $\nu = 0.1$ and $\nu=1.0$.

%%%%%%%%%%%%%%%%%%%%%%%%%%%%%%%%%%%%%%%%%%%%%%%%%%%%%%%%%%%%%%%%%%%%%%%%%%%%%%%
\section{Strongly nonlinear regime\label{sec:strongly:nonlinear}}

The ``strongly nonlinear" regime defined in Ref.~\cite{bazant2004} is characterized by significant $O(1)$ perturbations to the salt concentration in the quasi-neutral diffusion layers. The perturbations are driven by the uptake of an excess amount of salt $w = \epsilon\tilde w$ into the diffuse layer from a diffusion zone of width $\sqrt{\epsilon/\omega}$. This induces a local $O(\epsilon\tilde w / \sqrt{\epsilon/\omega})$ drop in the concentration, and therefore we expect the strongly nonlinear regime to start at $\sqrt{\epsilon\omega}\langle\tilde w\rangle = O(1)$. For $\epsilon = 0.001$ and $\omega = O(1)$, Fig.~\ref{fig:wavg} indicates this is reached for $V \approx 30$.

From a physical point of view, this regime of the model is novel and interesting in several ways. It predicts the possibility of  ``capacitive desalination" of the bulk solution by an ac voltage, which is a remarkable example of rectification by nonlinearity, since even strong ac voltages are normally assumed not to perturb the bulk solution, in the absence of Faradaic reactions. This phenomenon may have interesting applications in microfluidics, since ac voltages are often used to apply large electric fields without triggering reactions. Second, concentration gradients in the oscillating diffusion layers can be come large enough to cause nearly complete depletion of salt just outside the double layer, causing to lose its quasi-equilibrium structure. This situation  of ``transient limiting current" is analyzed in the next section, but first we describe strongly nonlinear dynamics without diffusion limitation.

\begin{figure}
\includegraphics[width=0.45\textwidth]{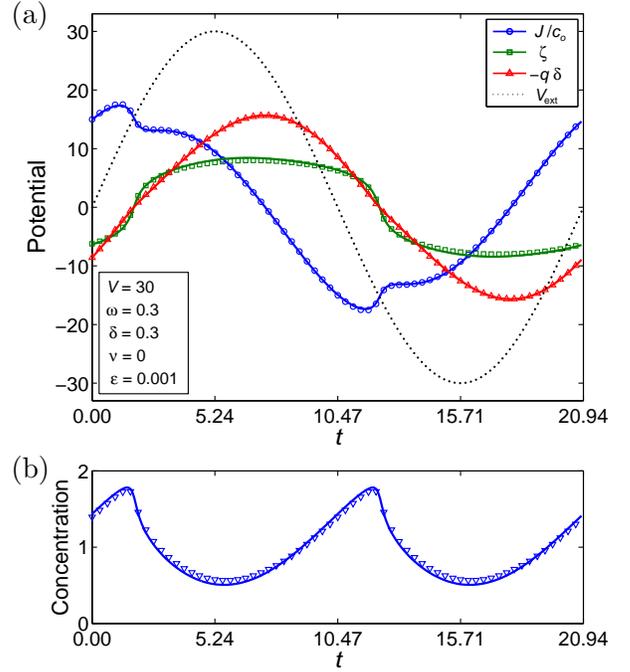}
\caption{(Color online) Strongly nonlinear response at $V = 30$, $\omega = 0.3$, $\delta = 0.3$, $\nu = 0$, and $\epsilon = 0.001$. (a) Distribution of the cell voltage, $V_\mathrm{ext} = \bar J/\bar c_o + \tilde\zeta - \tilde q\,\delta$ (dotted), onto the bulk electrolyte, $\bar J/\bar c_o$ (circles), diffuse layer, $\tilde\zeta$ (squares), and compact layer, $-\tilde q\,\delta$ (triangles). (b) Concentration $\hat c_s$ at the inner edge of the diffusion layer. Symbols show results from our full numerical solution of the PNP equations, while the solid lines are predictions of the (much simpler) uniformly valid asymptotic approximations, which are seen to be in excellent agreement.}
\label{fig:potentials30}
\end{figure}

%%%%%%%%%%%%%%%%%%%%%%%%%%%%%%%%%%%%%%%%%%%%%%%%%%%%%%%%%%%%%%%%%%%%%%%%%%%%%%%
\subsection{Dynamical response\label{sec:time:series:strongly}}

Figure~\ref{fig:potentials30} shows the strongly nonlinear dynamic response at at $V = 30$, $\omega = 0.3$, $\delta = 0.3$, $\nu = 0$, and $\epsilon = 0.001$. First off we note that the qualitative difference against the weakly nonlinear solution from Fig.~\ref{fig:signal:weakly}(b) is fairly small, even though the surface concentration $\hat c_s$ shows a significant variation. Quantitatively the largest difference is on the zeta potential, reaching 12\% relative difference between the weakly and strongly nonlinear models.
Perhaps this should not be too surprising: The surface concentration affects the double-layer charging dynamics only through the diffuse-layer charge-voltage relation, Eq.~\eqref{eq:excess:charge:steric}, and only in a square-root dependence. Moreover, at this voltage the diffuse-layer capacitance is large enough that the compact layer dominates the overall response. Hence for smaller values of $\delta$, we should see a more significant difference between the weakly and strongly nonlinear regimes. On the other hand, for $\nu\neq 0$ the double-layer voltage eventúally becomes dominated by the condensed phase of ions developing at the steric limit, which scales as $|\tilde\zeta|\approx \nu\tilde q^2/2$ \emph{independent} of $\hat c_s$.

%%%%%%%%%%%%%%%%%%%%%%%%%%%%%%%%%%%%%%%%%%%%%%%%%%%%%%%%%%%%%%%%%%%%%%%%%%%%%%%
\subsection{Numerical validation\label{sec:validation:strongly}}

In order to test our uniformly valid asymptotic approximations above in the strongly nonlinear regime, we compare the results to the full numerical solution of the PNP model from Fig.~\ref{fig:pnp30}: In Figure~\ref{fig:potentials30} the solid lines show the results from the asymptotic analysis, and symbols show corresponding output from the full PNP model, determined in the following way: The compact-layer voltage, $-\tilde q^\mathrm{PNP}\delta$, is given directly by Eq.~\eqref{eq:stern:bc}, the bulk current, $\bar J^\mathrm{PNP}$, and salt concentration, $\bar c_o^\mathrm{PNP}$, are evaluated at the center of the cell at $x = 0$, the diffuse-layer voltage, $\tilde\zeta^\mathrm{PNP}$, is computed as the potential drop from the electrode surface at $y = 0$ (i.e., $x=-1$) to a point immediately outside the diffuse layer, chosen (arbitrarily) at $y=3\epsilon$, and likewise the concentration $\hat c_s^\mathrm{PNP}$ is evaluated at $y = 3\epsilon$.

Overall, the agreement between the full PNP numerical solutions and the uniformly valid asymptotic approximations is excellent, in spite of the dramatic mathematical simplification at large voltages. The bulk current in the asymptotic model is slightly too small when $\hat c_s$ is maximal, and slightly too large when $\hat c_s$ is minimal, with a maximal relative error of 1\%, measured as $\max_t|\bar J-\bar J^\mathrm{PNP}|/\max_t|\bar J^\mathrm{PNP}|$.
This small discrepancy is primarily due to our neglect of the change in conductivity in the diffusion layer and the associated (small) excess voltage, cf. Eq.~\eqref{eq:excess:ohmic:field}. The compact-layer voltage agrees very well with the full numerical solution, whereas the diffuse-layer voltage $\tilde\zeta$ appears to be about 6\%{} too large.
However, the excess potential $\tilde\psi$ in the diffuse layer falls off exponentially at large $\tilde y$, cf. Eq.~\eqref{eq:gouy:chapman}, so measuring the diffuse-layer voltage in the PNP model from $\tilde y = 0$ to $\tilde y = 3$ we miss a (small) fraction of the ``true" result. Accounting for this, we find the relative error is only 2\%{}, mainly due to a phase lag between the two solutions.
The same arguments apply to the salt concentration $\hat c_s$ at the inner ``edge" of the diffusion layer: Using Eq.~\eqref{eq:boltzmann} to compute $\tilde c$ at $\tilde y = 3$ the agreement with the full numerical solution is accurate to within 1\%, against 4\% for the ``raw" $\hat c_s$ data in Fig.~\ref{fig:potentials30}(b).

%%%%%%%%%%%%%%%%%%%%%%%%%%%%%%%%%%%%%%%%%%%%%%%%%%%%%%%%%%%%%%%%%%%%%%%%%%%%%%%
\subsection{Local salt depletion\label{sec:depletion}}

\begin{figure}
\includegraphics[width=0.45\textwidth]{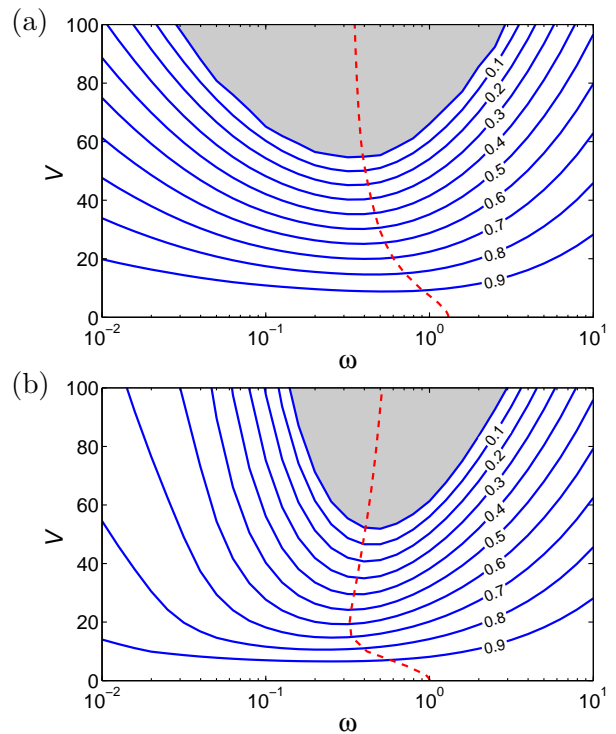}
\caption{(Color online) Contour plot of the minimal salt concentration $\min_t\hat c_s$ in the diffusion layer as a function of driving voltage and frequency for (a) $\delta = 0.3$, $\nu = 0$, and $\epsilon = 0.001$; (b) $\delta = 0$, $\nu = 0.01$, and $\epsilon = 0.001$. The dashed line marks the characteristic frequency $\omega_o$ and the shaded area marks a regime where $\min_t\hat c_s$ drops to zero and the double layer is driven out of quasi-equilibrium. (For the lowest frequencies in the figure, the $O(\sqrt{\epsilon/\omega})\approx 0.3$ diffusion layers extend across most of the bulk and no longer act as mathematical boundary layers.)}
\label{fig:cmin}
\end{figure}

%The strength of the oscillations in the local salt concentration in the diffusion layer depends on the driving frequency. We already argued that the perturbations scale as $O(\langle\tilde w\rangle/\sqrt{\epsilon\omega})$ when an excess amount $w = \epsilon\tilde w$ of neutral salt is taken up from a region of width $\sqrt{\epsilon/\omega}$. Hence we expect the salt depletion to become less significant at low frequency. On the other hand, the accumulation of salt is most significant when the system is driven at or below the characteristic frequency $\omega_o$, cf. Fig.~\ref{fig:wavg}. Overall we therefore expect the variations in the diffusion layer concentrations to be strongest for $\omega\approx\omega_o$. Fig.~\ref{fig:cmin} demonstrates this in detail, showing the minimal concentration $\min_t\{\hat c_s\}$ in the diffusion layer over a full period in time as a function of driving frequency and voltage. Fig.~\ref{fig:cmin}(a) shows the results for $\delta = 0.3$, $\nu = 0$, and $\epsilon = 0.001$, where it is clearly seen that the salt depletion is stronges for $\omega\approx\omega_o$.

In order to quantify the strength of the nonlinear response, we measure the minimal concentration in the diffusion layer over one period in time. For example, in Fig.~\ref{fig:potentials30}(b) the minimal concentration is about $\min_t\hat c_s \approx 0.5$, which is attained just after $t = 5.24$ and again after $t=15.71$. Figure~\ref{fig:cmin} shows the result for $\min_t\hat c_s$ as a function of driving frequency and voltage. Figure~\ref{fig:cmin}(a) shows the result for the GCS model with $\delta = 0.3$, $\nu = 0$, and $\epsilon = 0.001$, with at least two important points to note: First, at a given driving voltage, the salt depletion is most significant just around the characteristic frequency $\omega_o$, and second, at a given driving frequency $\min_t\hat c_s$ falls off roughly linearly with $V$, scaling as $1- O(V\sqrt{\epsilon\omega}/\delta)$ for $\omega\lesssim \omega_o$.

The frequency dependence can be understood as follows. Earlier we argued that when the diffuse layer absorbs neutral salt from a diffusion layer of width $\sqrt{\epsilon/\omega}$, the variations in $\hat c_s$ should scale as $\sqrt{\epsilon\omega}\langle\tilde w\rangle$, which explains why the salt depletion becomes less significant at low frequency. On the other hand, the double layer only gets fully charged when the system is driven below the characteristic frequency $\omega_o$, cf. Fig.~\ref{fig:wavg}, so that overall we should indeed expect to see the strongest salt depletion for $\omega\approx\omega_o$.

Another important feature of the strongly nonlinear regime is the possibility of transient diffusion limitation. This occurs when the voltage is sufficiently large, and the frequency sufficiently small, to temporarily, but  completely, deplete the salt concentration at the inner edge of the diffusion layer.   The shaded area in Fig.~\ref{fig:cmin}(a) marks the parameter range where $\min_t\hat c_s$ hits zero, and the quasi-equilibrium structure of the double layer breaks down. In this novel regime, we must revise our asymptotic analysis to produce uniformly valid approximations accounting for transient space charge formation. This is the subject of Sec.~\ref{sec:asymptotics:beyond} below.

Volume constraints can have a significant effect on the strongly nonlinear dynamics of our model problem.
Figure~\ref{fig:cmin}(b) shows the corresponding results for Bikerman's model with $\delta = 0$, $\nu = 0.01$, and $\epsilon = 0.001$. Again, the salt depletion is strongest when the system is driven around the characteristic frequency, although this has a different dependence on voltage, as noted above. Further it is clear that when steric exclusion sets in and the diffuse-layer capacitance decreases, the salt depletion becomes much less significant, especially at low frequency. This effect was noted by Kilic et al~\cite{kilic2007a,kilic2007b} for the response to a sudden dc voltage, but its influence on strongly nonlinear ac response is more complicated. Steric effects make the shaded area of transient diffusion limitation span a narrower range of frequencies, compared to the GCS model.  However, in both models, the shaded area starts at roughly the same voltage for the characteristic frequency.

%%%%%%%%%%%%%%%%%%%%%%%%%%%%%%%%%%%%%%%%%%%%%%%%%%%%%%%%%%%%%%%%%%%%%%%%%%%%%%%
\section{Breakdown of quasi-equilibrium double-layer structure\label{sec:asymptotics:beyond}}

As we have seen in Figure~\ref{fig:cmin}, when the driving voltage is increased, the salt depletion in the diffusion layer becomes more and more pronounced, and at some point the minimal concentration can even drop to zero (within the shaded area). At that point, the quasi-equilibrium structure of the double layer breaks down: The chemical potential diverges, and the effective width of the diffuse layer grows like $O(1/\sqrt{\hat c_s})$. Likewise, the quasi-electroneutral solution in the diffusion layer breaks down when the concentration approaches zero: The leading order charge density in the diffusion layer can be evaluated from the Poisson equation, cf. Eq.~\eqref{eq:diffusion:charge},
\[
\hat\rho = -\epsilon^{3/2}\frac{\bar J\partial_{\hat y}\hat c}{\hat c^2}.
\]
At large voltage the flux into the double layer is dominated by uptake of counterions since there are no more coions to expel, so that at the inner edge we have $|\bar J| \approx \partial_{\hat y}\hat c/\sqrt{\epsilon}$ and $|\hat\rho| \approx \epsilon^2\bar J^2/\hat c^2$. Quasi-electroneutrality in the diffusion layer remains a good approximation only as long as $\hat c \gg |\hat\rho|$ or
\beq
\hat c_s \gg |\epsilon\bar J|^{2/3}.
\eeq
The breakdown of electroneutrality and concommitant expansion of the double layer into a non-equilibrium structure due to transient diffusion limitation, in the absence of any normal flux of ions at the electrodes, is a novel prediction of our model which we analyze in detail in this section.

%%%%%%%%%%%%%%%%%%%%%%%%%%%%%%%%%%%%%%%%%%%%%%%%%%%%%%%%%%%%%%%%%%%%%%%%%%%%%%%
\subsection{Nonequilibrium double layer\label{sec:nonequilibrium:dl}}

The breakdown of quasi-equilibrium in the double layer and of quasi-electroneutrality in the bulk region is well known for electrochemical cells driven at a dc Faradaic current approaching the classical ``limiting" current~\cite{bazant2005,chu2005}. At the limiting dc current, the double layer acquires a {\it steady} non-equilibrium structure and expands in dimensionless width from $O(\epsilon)$ to $O(\epsilon^{2/3})$, as first described by Smyrl and Newman~\cite{smyrl1967}. Rubinstein and Shtilman later showed that a ``space charge" region completely depleted of coions can develop at an electrode or ion exchange membrane when driven \emph{above} the diffusion-limited current~\cite{rubinstein1979}. In this regime one can identify three sublayers within the nonequilibrium double layer, namely~\cite{rubinstein2001,rubinstein2005,chu2005}
\begin{itemize}
\item
An inner quasi-equilibrium layer of width $O(\epsilon)$ at the electrode surface.
\item
An extended ``space-charge" layer of width $y_o> O(\epsilon^{2/3})$ that is completely depleted of coions.
\item
A ``Smyrl-Newman" transition layer of width $O(\epsilon^{2/3})$ around $y = y_o$ connecting the space-charge layer to the quasi-electroneutral diffusion layer.
\end{itemize}
It is exactly the same nested boundary-layer structure that we see here develop in a cell driven dynamically at very large voltage even though the electrodes are blocking with no reactions taking place, and thus no normal flux of ions into or out of the cell. Instead, the double layer is driven out of equilibrium purely by nonlinear electrochemical relaxation within the cell, as counterions are absorbed into the double layer so quickly and in such large numbers that bulk diffusion becomes {\it temporarily} rate limiting, within each ac period. We now develop uniformly valid asymptotic approximations for this novel regime.

%%%%%%%%%%%%%%%%%%%%%%%%%%%%%%%%%%%%%%%%%%%%%%%%%%%%%%%%%%%%%%%%%%%%%%%%%%%%%%%
\subsubsection{Space-charge layer}

When a negative voltage is applied on the left electrode, the space-charge layer developing is one completely depleted of anions, $\breve c_-=0$ (denoting variables by a breve accent), while the cation concentration is nonzero, $\breve c_+>0$, but small. The ion transport is completely dominated by migration, and the flux is determined by the current fed into the boundary layers from the bulk
\beq
\frac{1}{2}\breve c_+\partial_y\breve\phi = |\bar J(t)|.
\eeq
Substituting the Poisson equation $-\epsilon^2\partial_y^2\breve\phi = \frac{1}{2}\breve c_+$ and integrating, we get the leading order field in the space-charge layer
\beq[space:charge:field]
\partial_y\breve\phi = \frac{1}{\epsilon}\sqrt{2|\bar J|(y_o-y)},
\eeq
where the integration constant $y_o(t)$ is born positive and equal to the width of the space-charge layer. The (small) charge density due to the counterions in the cationic space-charge layer is found by differentiation,
\beq[space:charge:density]
\breve\rho = \frac{\breve c_+}{2} = -\epsilon^2\partial_y^2\breve\phi = \frac{\epsilon}{2}\sqrt{\frac{2|\bar J|}{y_o-y}},
\eeq
and the leading order potential drop across the layer by integration
\beq[space:charge:potential]
\breve\Phi = \breve\phi(0) - \breve\phi(y_o) = -\frac{2}{3\epsilon}\sqrt{2|\bar J|}y_o^{3/2}.
\eeq
The analysis of the opposite case, where a positive voltage is applied on the electrode and a space-charge layer completely depleted of cations develops, is fully similar. %, but with opposite sign of $\breve\Phi$, $\breve\rho$, and $\partial_y\breve\phi$.

%%%%%%%%%%%%%%%%%%%%%%%%%%%%%%%%%%%%%%%%%%%%%%%%%%%%%%%%%%%%%%%%%%%%%%%%%%%%%%%
\subsubsection{Inner diffuse layer\label{sec:neq:inner:diffuse}}

Within an $O(\epsilon)$ distance from the electrode surface, the counterions remain in quasi-equilibrium with a constant electrochemical potential at leading order~\cite{rubinstein2001}, i.e., for a space-charge layer completely depleted of anions
\beq[inner:chemical:potential]
\tilde\mu_+ = \tilde\phi + \log\tilde c_+ - \log(1-\nu\tilde c_+/2) = const.
\eeq
from which the cation distribution is
\beq
\tilde c_+ = \frac{1}{\frac{\nu}{2}+e^{\tilde\phi-\tilde\mu_+}}.
\eeq
Substituting into the Poisson equation and integrating, we find
\beq[inner:field:mpb]
\partial_{\tilde y}\tilde\phi = \sqrt{\kappa^2 + \frac{2}{\nu}\log\Big[1 + \frac{\nu}{2} e^{\tilde\mu_+ - \tilde\phi}\Big]},
\eeq
where the integration constant $\kappa$ is fixed as $\kappa = \sqrt{2|\bar J|y_o} = \epsilon\partial_y\breve\phi(0)$ to match the field in the space-charge layer. The solution for the potential can be expressed in integral form as
\beq
\tilde y = \int_{\tilde\phi(0)}^{\tilde\phi(\tilde y)} \frac{\id\tilde\phi'}{\sqrt{\kappa^2 + \frac{2}{\nu}\log\Big[1 + \frac{\nu}{2} e^{\tilde\mu_+ - \tilde\phi'}\Big]}}.
\eeq
The difficulty is, however, that we cannot determine the chemical potential $\tilde\mu_+$ by matching with the space-charge layer because $\tilde\phi\to\infty$ and $\tilde c_+\to 0$ for $\tilde y \to\infty$.

In the dilute limit $\tilde c_+\ll 2/\nu$ the solution can be expressed in \emph{closed} form~\cite{chu2005,zaltzman2007}. Rewriting in terms of the excess potential $\tilde\psi = \tilde\phi - \breve\phi$ we obtain
\beq[inner:excess:potential]
\tilde\psi = 2\log\big[1-\big(1-e^{\tilde\zeta/2}\big)e^{-\kappa\tilde y}\big],
\eeq
where $\tilde\zeta$ is determined by the total charge $\tilde q = \partial_{\tilde y}\tilde\phi(0) = \partial_{\tilde y}\tilde\psi(0) + \kappa$ accumulated in the non-equilibrium double layer,
\beq[voltage:charge:dilute]
\tilde\zeta = -2\log\bigg[\frac{1}{2} + \frac{\tilde q}{2\kappa}\bigg].
\eeq
Substituting Eq.~\eqref{eq:inner:excess:potential} and $\tilde c_+ = -2\partial_{\tilde y}^2\tilde\psi$ into Eq.~\eqref{eq:inner:chemical:potential} we find
\begin{align}
\tilde\mu_+ &= \breve\phi(0) + 2\log(2\kappa) + \log\big(1-e^{\tilde\zeta/2}\big) \label{eq:dilute:chemical:potential}\\
 &= \breve\phi(0) + 2\log(2\kappa) + \log\bigg(\frac{\tilde q-\kappa}{\tilde q+\kappa}\bigg).
\end{align}
We note that for $\tilde q \gg \kappa$ the chemical potential approaches a level $\tilde\mu_+ = \breve\phi(0) + 2\log(2\kappa)$ that is independent of $\tilde\zeta$ or $\tilde q$ and determined only by the matching field from the space-charge layer.

Now, provided $\kappa$ is much smaller than the field at the onset of steric exclusion, i.e., $\kappa\ll\sqrt{2/\nu}$, we can use Eq.~\eqref{eq:dilute:chemical:potential} for the chemical potential also for the MPB problem. Substituting into Eq.~\eqref{eq:inner:field:mpb} we then get
\beq
\tilde q = \sqrt{\kappa^2 + \frac{2}{\nu}\log\big[1 + 2\nu\kappa^2\big(1-e^{\tilde\zeta/2}\big) e^{-\tilde\zeta}\big]},
\eeq
and, finally, solving for $\tilde\zeta$ and using $\kappa\ll\sqrt{2/\nu}$
\beq[voltage:charge:steric]
\tilde\zeta \simeq -2\log\bigg[\frac{1}{2} + \frac{1}{2\kappa}\sqrt{2\big(e^{\nu\tilde q^2/2}-1\big)/\nu}\bigg].
\eeq

%%%%%%%%%%%%%%%%%%%%%%%%%%%%%%%%%%%%%%%%%%%%%%%%%%%%%%%%%%%%%%%%%%%%%%%%%%%%%%%
\subsubsection{Transition layer}

The solution in the space-charge layer cannot be matched directly to the quasi-electroneutral diffusion layer: The leading order field from the space-charge layer is $O(\epsilon^{-1})$ but vanishes at $y = y_o$, whereas the field in the diffusion layer is $O(1)$ but diverges like $-\bar J/\hat c \approx 1/(y-y_o)$, cf. Eqs.~\eqref{eq:excess:ohmic:field} and \eqref{eq:space:charge:field}.

In the transition zone for $|y-y_o|\leq O(\epsilon^{2/3})$ the field has a unique profile that can be expressed as~\cite{rubinstein2001}
\beq
\partial_y\phi = \frac{\bar J P(z)}{|\epsilon\bar J|^{2/3}},
\eeq
where $P(z)$ is a Painlev\'e transcendent of the second kind and $z$ is a rescaled spatial variable given by
\beq
z = \frac{|\bar J|(y-y_o)+\hat c_s}{|\epsilon\bar J|^{2/3}}.
\eeq
See the Appendix for more details and a plot of $P(z)$.

The voltage on this narrow layer is negligible in the overall cell response, but the solution does allow us to understand better both the spatial transition, and the transition in time from quasi- to nonequilibrium structure.

%%%%%%%%%%%%%%%%%%%%%%%%%%%%%%%%%%%%%%%%%%%%%%%%%%%%%%%%%%%%%%%%%%%%%%%%%%%%%%%
\subsubsection{Transition from quasi- to nonequilibrium}

One challenging aspect in modelling the dynamics of the system is that (unlike the steady-state dc case analyzed in all prior work) it is not sufficient to have valid solutions for quasi-equilibrium with $\hat c_s \gg \epsilon^{2/3}$ and nonequilibrium with $y_o\gg\epsilon^{2/3}$. The dynamical response passes back and forth from quasi-equilibrium to nonequilibrium, but the previous analysis of the inner diffuse layer from Secs.~\ref{sec:equilibrium:dl} and~\ref{sec:neq:inner:diffuse} breaks down in the transition regime from $\hat c_s = O(\epsilon^{2/3})$ to $y_o = O(\epsilon^{2/3})$, leaving the diffuse-layer voltage from both Eqs.~\eqref{eq:excess:voltage:steric} and~\eqref{eq:voltage:charge:steric} divergent for $\hat c_s\to 0$ and $y_o\to 0$, respectively, at fixed $\tilde q$. This is problematic since the charge-voltage relation plays a central role in our dynamical model.

In order to resolve this, we have developed an approximate solution of the standard PNP equations for the inner diffuse layer, that is uniformly valid both in quasi-equilibrium, nonequilibrium, and across the transition regime. Essentially, our approximation amounts to assuming constant, rather than variable, coefficients in the equation for the \emph{excess} field in the boundary layer from the Painlev\'e II problem; see the Appendix for technical details. Then we obtain
\begin{align}
\tilde\psi &= 4\tanh^{-1}\Bigg[\frac{\tanh(\tilde\zeta/4)e^{-\kappa\tilde y}}{1-\frac{\breve\kappa}{\kappa}\tanh(\tilde\zeta/4)(1-e^{-\kappa\tilde y})}\Bigg], \label{eq:excess:potential:general}
\intertext{where}
\kappa &= |\epsilon\bar J|^{1/3}\sqrt{3P(-z_o)^2/2-z_o}, \\
\breve\kappa &= |\epsilon\bar J|^{1/3}\mathrm{sign}(\bar J)P(-z_o), \\
z_o &= \big(|\bar J|y_o-\hat c_s\big)/|\epsilon\bar J|^{2/3},
\intertext{and the charge-voltage relation is}
\tilde\zeta &= -2\log\Bigg(\sqrt{\bigg[\frac{\tilde q+\breve\kappa}{2(\kappa+\breve\kappa)}\bigg]^2 + \frac{\kappa-\breve\kappa}{\kappa+\breve\kappa}} + \frac{\tilde q+\breve\kappa}{2(\kappa+\breve\kappa)}\Bigg). \label{eq:voltage:charge:general}
\end{align}
The asymptotics of $P(z)$ are such that in the quasi-equilibrium limit, where $\hat c_s\gg\epsilon^{2/3}$ and $z_o\ll-1$, we have $P(-z_o) \approx 1/z_o$, $\kappa\approx\sqrt{\hat c_s}$, $\breve\kappa\approx 0$, and we recover the standard Gouy--Chapman solution. In the nonequilibrium limit, where $y_o\gg\epsilon^{2/3}$ and $z_o\gg1$, we have $P(-z_o)\approx -\sqrt{2z_o}$, $\kappa\approx|\breve\kappa|\approx\sqrt{2|\bar J|y_o}$, and we recover the result from Eq.~\eqref{eq:voltage:charge:dilute}.

Extending this analysis to account for steric exclusion does not affect the space-charge layer, since the concentration there is low. It does, however, affect the solution in the inner diffuse layer. We are not able to produce an explicit analytical solution like Eq.~\eqref{eq:excess:potential:general} to the MPNP equations. But, noting that the MPB charge-voltage relations both quasi-equilibrium and nonequilibrium cases, Eqs.~\eqref{eq:excess:voltage:steric} and~\eqref{eq:voltage:charge:steric} are obtained by substituting $\tilde q$ in the corresponding PB results with $\mathrm{sign}(\tilde q)\sqrt{2(e^{\nu\tilde q^2/2}-1)/\nu}$. Extrapolating this observation to the general form in Eq.~\eqref{eq:voltage:charge:general}, we argue that the general MPB charge voltage relation is simply obtained by replacing $\tilde q$ with $\mathrm{sign}(\tilde q)\sqrt{2(e^{\nu\tilde q^2/2}-1)/\nu}$ in Eq.~\eqref{eq:voltage:charge:general}. At least this form reduces to Eq.~\eqref{eq:voltage:charge:general} for $\tilde q^2\ll2/\nu$, to Eqs.~\eqref{eq:excess:voltage:steric} and~\eqref{eq:voltage:charge:steric} for $\hat c_s\gg\epsilon^{2/3}$ and $y_o\gg\epsilon^{2/3}$, respectively, and to $\tilde\zeta\approx-\mathrm{sign}(\tilde q)\nu\tilde q^2$ for $\tilde q^2\gg2/\nu$.

%%%%%%%%%%%%%%%%%%%%%%%%%%%%%%%%%%%%%%%%%%%%%%%%%%%%%%%%%%%%%%%%%%%%%%%%%%%%%%%
\subsection{Modified diffusion layer\label{sec:diffusion:layer:revisited}}

The width $y_o(t)$ of the space-charge layer is equal to the width of the region of complete salt depletion in the diffusion layer. In the quasi-electroneutral part we still need to solve a simple diffusion problem
\[
\partial_t\hat c = \partial_{\hat y}^2\hat c,
\]
however, now it is to be solved on the dynamically changing interval $\hat y\in[\hat y_o(t),\infty)$, where $\hat y_o = y_o/\sqrt{\epsilon}$ is determined such that
\begin{alignat}{2}
\hat y_o(t) &= 0 \quad\mbox{for}\quad &\hat c(0,t) &> 0, \\
\hat c(\hat y_o(t),t) &= 0 \quad\mbox{for}\quad &\hat y_o(t) &> 0. \label{eq:diffusion:edge:def}
\end{alignat}
At the inner edge of the diffusion layer, the boundary condition is still obtained by matching with the salt flux out of the double layer
\beq[diffusion:bc:edge]
-\frac{1}{\sqrt{\epsilon}}\lim_{\hat y\to\hat y_o^+}\partial_{\hat y}\hat c = \lim_{\hat y\to\hat y_o^+}\hat F \equiv \tilde F_o(t).
\eeq
The problem can be solved using the method of images
\begin{align}
\hat c =&\, \bar c_o + \sqrt{\epsilon}\int_0^T\frac{1}{2}\big[G_\omega(|\hat y-\hat y_o(t')|,t-t') \nonumber \\
 & +G_\omega(\hat y+\hat y_o(t'),t-t')\big]\tilde F_o(t')\,\id t' , \label{eq:convolution:noneq}
\end{align}
where the injection of flux at both $\pm\hat y_o(t')$ makes the origin act like a reflecting boundary, and the Green's function $G_\omega(\hat y,t)$ is defined in Eq.~\eqref{eq:periodic:kernel}. The definition of $\hat y_o$ through Eq.~\eqref{eq:diffusion:edge:def} ensures that $\hat c = 0$ and $\partial_{\hat y}\hat c = 0$ for $\hat y < \hat y_o(t)$, and hence, using $\mp\partial_{\hat y}G_\omega(0^\pm,t) = \delta^+(t)$, the boundary condition \eqref{eq:diffusion:bc:edge} is indeed satisfied.

Interestingly, the modified diffusion layer does have a time-average excess salt content relative to the bulk: Substituting Eq.~\eqref{eq:periodic:kernel} in Eq.~\eqref{eq:convolution:noneq} we find that
\beq
\langle\hat c\rangle = \bar c_o - \frac{\sqrt{\epsilon}}{2}\big<\big(|\hat y-\hat y_o|+|\hat y+\hat y_o|\big)\tilde F_o\big>,
\eeq
and hence the time-average excess amount of salt contained is
\beq[diffusion:layer:excess:salt]
\sqrt{\epsilon}\langle\hat w\rangle = \sqrt{\epsilon}\int_0^\infty\langle\hat c\rangle - \bar c\,\id\hat y = -\frac{\epsilon}{2}\langle\hat y_o\tilde F_o\rangle,
\eeq
where we used that $\langle\tilde F_o\rangle = -\langle\partial_t\tilde w\rangle = 0$~\footnote{The result \eqref{eq:diffusion:layer:excess:salt} holds more generally when $\langle\tilde F_o\rangle\neq0$, in steady-state or for slow dynamics on the bulk diffusion time scale $\bar t = \epsilon t$, because $\nbf\cdot\nablabf\bar c = -\nbf\cdot\bar{\Fbf} = -\langle\tilde F_o\rangle$, cf. Eq.~\eqref{eq:salt:flux:bc:2D}.}.
Note that since $\tilde F_o$ is generally negative when $\hat y_o$ is nonzero, the excess salt content will be positive.

Since the local conductivity goes to zero when $\hat c$ vanishes at the inner edge, one might worry that the potential drop across the diffusion layer could be large. However, because the quasi-electroneutral solution breaks down for $\hat c \lesssim |\epsilon \bar J|^{2/3}$, we find that the overall ohmic potential drop across the quasi-electroneutral diffusion layer is limited to $O\big(\log|\epsilon\bar J|^{2/3}\big)$, which remains a small perturbation to that across the bulk electrolyte.

\begin{figure}
\includegraphics[width=0.45\textwidth]{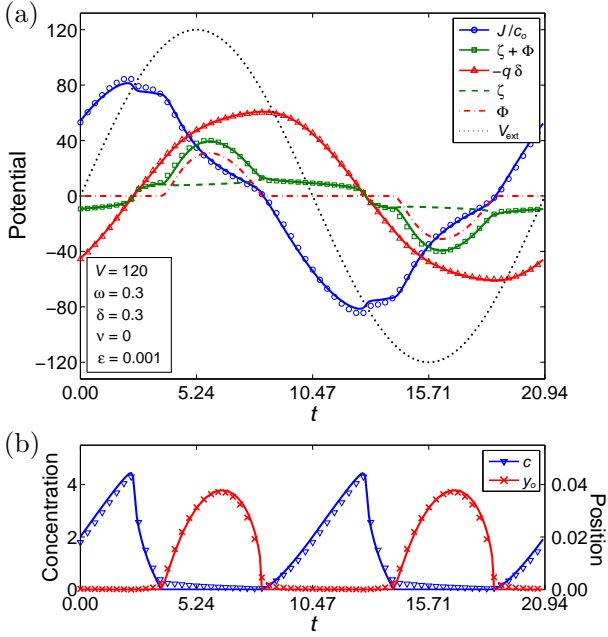}
\caption{(Color online) Strongly nonlinear response at $V = 120$, $\omega = 0.3$, $\delta = 0.3$, $\nu = 0$, and $\epsilon = 0.001$. (a) Distribution of the cell voltage, $V_\mathrm{ext} = \bar J/\bar c_o + \breve\Phi + \tilde\zeta - \tilde q\,\delta$ (dotted), onto the bulk electrolyte, $\bar J/\bar c_o$ (circles), diffuse layer (dashed), space-charge layer (dash-dot), and compact layer, $-\tilde q\,\delta$ (triangles); also shown is the sum of the space-charge and diffuse-layer voltages, $\tilde\zeta+\breve\Phi$ (squares). (b) Concentration $\hat c_s$ at the inner edge of the diffusion layer (triangles) and extent $y_o$ of space-charge layer (crosses). Solid and broken lines show results from our asymptotic model, and symbols show the full numerical solution of the PNP equations.}
\label{fig:potentials120}
\end{figure}

%%%%%%%%%%%%%%%%%%%%%%%%%%%%%%%%%%%%%%%%%%%%%%%%%%%%%%%%%%%%%%%%%%%%%%%%%%%%%%%
\subsection{Model summary\label{sec:neq:model:summary}}

Summarizing the model for the leading order dynamic out-of-quasi-equilibrium response, the charging of the double-layer at the leftmost electrode is governed by
\begin{align}
V_\mathrm{ext} &= \bar J / \bar c_o + \breve\Phi + \tilde\zeta - \tilde q\,\delta, \\
\partial_t\tilde q &= - \bar J, \\
\tilde w &= \sqrt{\tilde q^2 + 4\hat c_s} - \sqrt{4\hat c_s}, \\
\partial_t\tilde w &= -\tilde F_o,
\end{align}
with the diffuse- and space-charge layer voltages $\tilde\zeta$ and $\breve\Phi$ given by
\begin{align}
\tilde\zeta &= -2\log\Bigg(\sqrt{\bigg[\frac{\mathcal{Q}+\breve\kappa}{2(\kappa+\breve\kappa)}\bigg]^2 + \frac{\kappa-\breve\kappa}{\kappa+\breve\kappa}} + \frac{\mathcal{Q}+\breve\kappa}{2(\kappa+\breve\kappa)}\Bigg), \\
\breve\Phi &= \mathrm{sign}(\bar J)\frac{2}{3\epsilon}\sqrt{2|\bar J|}y_o^{3/2}.
\end{align}
Here $\mathcal{Q} = \mathrm{sign}(\tilde q)\sqrt{2[\exp(\nu\tilde q^2/2)-1]/\nu}$ in the MPNP model, reducing to $\mathcal{Q} = \tilde q$ in the standard PNP model. The salt concentration $\hat c_s$ at the inner edge of the diffusion layer is determined from
\begin{align}
\hat c_s &= \bar c_o + \sqrt{\epsilon}\frac{1}{T}\int_0^T\frac{1}{2}\Big[G_\omega(|\hat y_o(t)-\hat y_o(t')|,t-t') \nonumber \\
 & +G_\omega(\hat y_o(t)+\hat y_o(t'),t-t')\Big]\tilde F_o(t')\,\id t' ,
\end{align}
and the width $y_o = \sqrt{\epsilon}\hat y_o$ of the space-charge layer from
\beq
\hat c_s\,y_o = 0, \ \hat c_s \geq 0, \ y_o\geq 0.
\eeq
The parameters $\kappa$ and $\breve\kappa$ are given by
\begin{align}
\kappa &= |\epsilon\bar J|^{2/3}\sqrt{3P(-z_o)^2/2-z_o}, \\
\breve\kappa &= \mathrm{sign}(\bar J)|\epsilon\bar J|^{2/3}P(-z_o),
\end{align}
where $P(z)$ is the Painlev\'e transcendent and $z_o = (|\bar J|y_o-\hat c_s)/|\epsilon\bar J|^{2/3}$ is the rescaled position of the transition layer relative to the electrode. Finally, the bulk concentration is again determined by imposing global conservation of salt in the cell to get
\beq
\bar{c}_o = 1 + \sqrt{\epsilon}\langle y_o\tilde F_o\rangle/2 - \epsilon\langle\tilde w\rangle.
\eeq

We solve this problem numerically using a timestepping algorithm. The major difficulty is to determine $y_o(t)$ in a self-consistent way, which we achieve using a bisection algorithm. More details are given in Ref.~\cite{epaps}.

%%%%%%%%%%%%%%%%%%%%%%%%%%%%%%%%%%%%%%%%%%%%%%%%%%%%%%%%%%%%%%%%%%%%%%%%%%%%%%%
\subsection{Dynamical response\label{sec:dynamic:response:neq}}

Figure~\ref{fig:potentials120} shows the strongly nonlinear dynamic response at $V = 120$, $\omega = 0.3$, $\delta = 0.3$, $\nu = 0$, and $\epsilon = 0.001$. In particular we note in Fig.~\ref{fig:potentials120}(b) the appearance of a transient space-charge layer extending to a width of $y_o\lesssim 0.04$, with an associated voltage $\breve\Phi$ in Fig.~\ref{fig:potentials120}(a) that induces a visible drop in the bulk current $\bar J$.

In order to validate to asymptotic analysis we compare the results to the full numerical solution from Fig.~\ref{fig:pnp120}, as shown with symbols in Fig.~\ref{fig:potentials120}. Like in Sec.~\ref{sec:validation:strongly}, the compact layer voltage $-\tilde q^\mathrm{PNP}\delta$ is given directly by Eq.~\eqref{eq:stern:bc}, and the bulk current $\bar J^\mathrm{PNP}$ and salt concentration $\bar c_o^\mathrm{PNP}$ are evaluated at the center of the cell. The width $y_o^\mathrm{PNP}$ of the space-charge layer is taken (arbitrarily) as the largest region where either the cation or anion concentration drops below $\frac{1}{2}|\epsilon\bar J|^{2/3}$, and, finally, the concentration $\hat c_s^\mathrm{PNP}$ and the overall voltage $(\tilde\zeta+\breve\Phi)^\mathrm{PNP}$ across the diffuse and space-charge layers are evaluated at the position $y_o^\mathrm{PNP}+3\epsilon$.

The agreement between the asymptotic approximation and the full numerical solution is rather good, although the bulk current $\bar J$ is visibly somewhat too low (by $\approx$ 5\%) when $\hat c_s$ is large and too high when the $\hat c_s$ is small. As before in Fig.~\ref{fig:potentials30}, part of the discrepancy on $\tilde\zeta$, $\breve\Phi$, and $\hat c_s$ is due to the difficulty with defining the diffusion-layer inner ``edge" on the full numerical solution, so it may be more appropriate to compare the full spatial profiles. We proceed to do that in the following section.

\begin{figure*}
\includegraphics[width=0.8\textwidth]{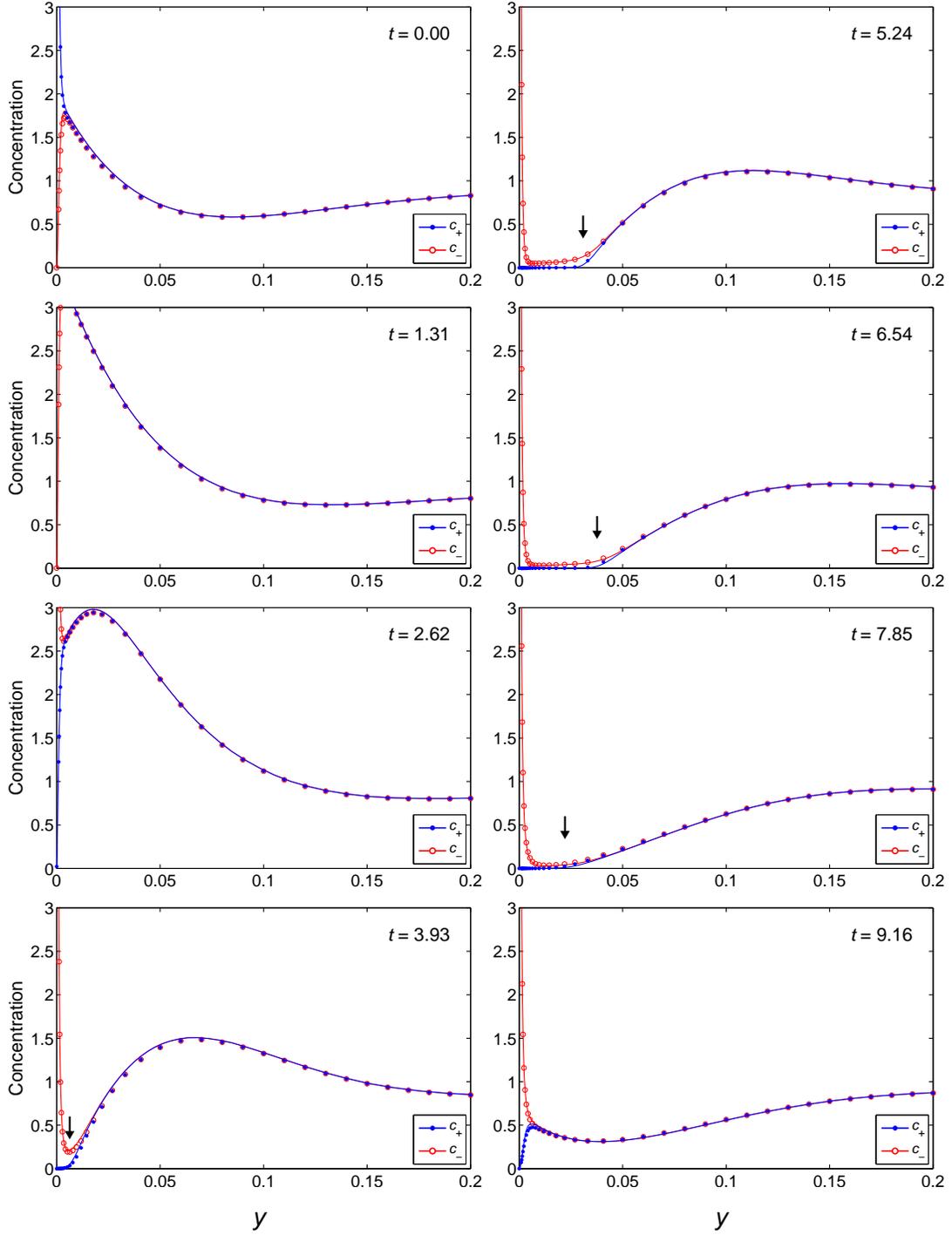}
\caption{(Color online) Concentration profiles at $V = 120$, $\omega = 0.3$, $\delta = 0.3$, $\nu = 0$, and $\epsilon = 0.001$; ``frames" cover one-half period in time starting from $t=0$. Open and filled circles show the cation and anion concentration, respectively, according to the full numerical solution from Fig.~\ref{fig:pnp120}, and solid lines show uniformly valid approximations based on the asymptotic analysis, with very good qualitative and quantitative agreement. Black arrows mark the extent $y_o(t)$ of the space-charge layer (if any) according to the asymptotic model.}
\label{fig:profiles120}
\end{figure*}

\begin{figure*}
\includegraphics[width=0.9\textwidth]{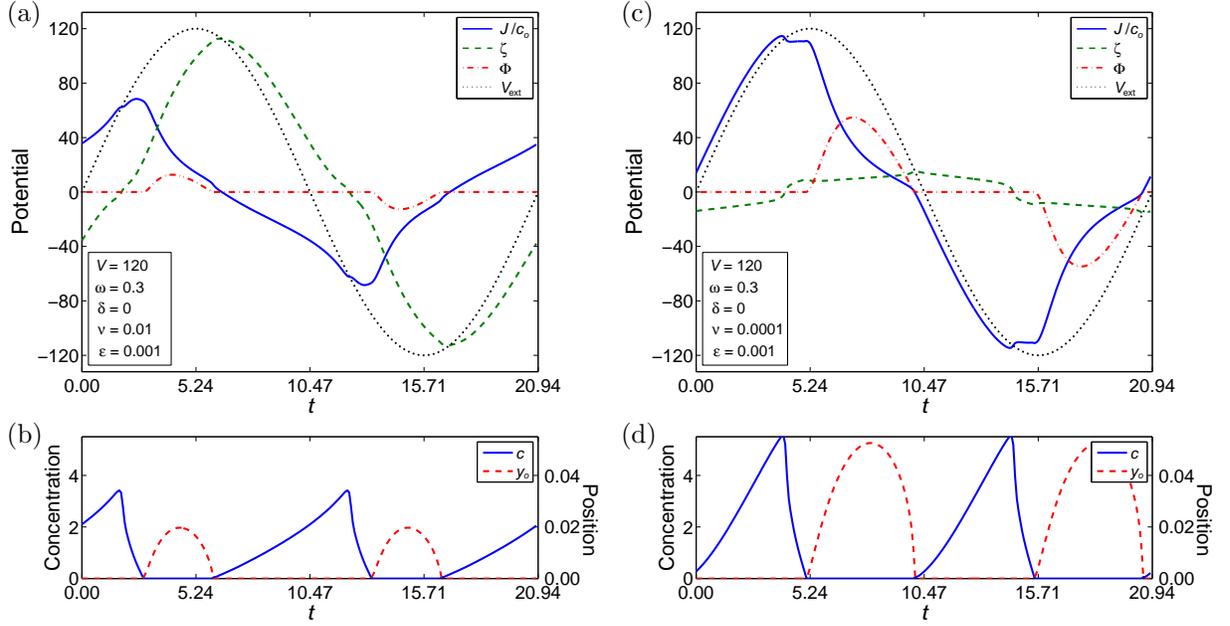}
\caption{(Color online) Strongly nonlinear response for MPB model with relatively large and small bulk ion volume fraction, $\nu = 0.01$ and 0.0001, respectively, at $V = 120$, $\omega = 0.3$, $\delta = 0$, and $\epsilon = 0.001$. (a) and (c) Distribution of the cell voltage, $V_\mathrm{ext} = \bar J/\bar c_o + \breve\Phi + \tilde\zeta$ (dotted), onto the bulk electrolyte, $\bar J/\bar c_o$ (solid), diffuse layer (dashed), and space-charge layer (dash-dot). (b) and (d) Concentration $\hat c_s$ at the inner edge of the diffusion layer (solid) and extent $y_o$ of space-charge layer (dashed).}
\label{fig:potentials120nu}
\end{figure*}

%%%%%%%%%%%%%%%%%%%%%%%%%%%%%%%%%%%%%%%%%%%%%%%%%%%%%%%%%%%%%%%%%%%%%%%%%%%%%%%
\subsection{Uniformly valid approximations\label{sec:uniformly:valid:approx}}

So far we have been focusing on integral quantities such as the total charge and voltage, but the asymptotic analysis also predict the full spatial profiles for the potential and ion concentrations in the cell: Uniformly valid approximations in space are constructed by adding the inner and outer approximations and subtracting the overlaps~\cite{bazant2004,kilic2007b}. In the absense of a space-charge layer the ion concentrations are given by
\begin{align}
c_\pm(x,t) &= \left[\tilde c_\pm\left(\frac{1+x}{\epsilon},t\right)-\hat c_s\right] + \hat c\left(\frac{1+x}{\sqrt{\epsilon}},t\right) - \bar c_o \nonumber \\
&+ \hat c\left(\frac{1-x}{\sqrt{\epsilon}},t\right) + \left[\tilde c_\mp\left(\frac{1-x}{\epsilon},t\right)-\hat c_s\right]. \label{eq:uniformly:valid:qeq}
\end{align}
In the presense of a space-charge layer, the challenge is to tie up the $O(\epsilon)$ but divergent counterion concentration in the space-charge layer, cf. Eq.~\eqref{eq:space:charge:density}, with the $O(1)$ but vanishing concentration in the diffusion layer at $y = y_o$. The key is to employ the solution from the Smyrl-Newman transition layer: In the transition to the space-charge layer on the left electrode, the concentration can be written as
\beq[transition:conc:1]
c_\pm = |\epsilon\bar J|^{2/3}(z + P^2/2 \mp \mathrm{sign}(\bar J)\partial_zP),
\eeq
as discussed in the Appendix. For $y\gg y_o$ this reduces to $c_+ = c_- = (y-y_o)|\bar J|$, matching the flux $\partial_{\hat y}\hat c/\sqrt{\epsilon} \approx |\bar J|$ at the inner edge of the diffusion layer, whereas for $y\ll y_o$ the space-charge density Eq.~\eqref{eq:space:charge:density} is recovered. It is convenient to rewrite Eq.~\eqref{eq:transition:conc:1} as $c_\pm = \hat c + \breve\varsigma_\pm$ where $\breve\varsigma_\pm = c_\pm - \hat c$ is the \emph{excess} concentration in the space-charge and transition layers relative to the diffusion layer, given by
\beq[transition:conc:2]
\breve\varsigma_\pm = |\epsilon\bar J|^{2/3}(\min\{z,0\} + P^2/2 \mp \mathrm{sign}(\bar J)\partial_zP).
\eeq
In line with this, the excess concentration in the inner diffuse layer can be expressed as
\begin{align}
\tilde\varsigma_\pm &= |\epsilon\bar J|^{2/3}(\tilde R^2/2 + P\tilde R \mp \mathrm{sign}(\bar J)\partial_z\tilde R) \nonumber \\
 &= (\partial_{\tilde y}\tilde\psi)^2/2 + \breve\kappa\partial_{\tilde y}\tilde\psi \mp \partial_{\tilde y}^2\tilde\psi
\end{align}
where $\tilde R = \mathrm{sign}(\bar J)\partial_z\tilde\psi = \mathrm{sign}(\bar J)\partial_{\tilde y}\tilde\psi/|\epsilon\bar J|^{1/3}$ is the (rescaled) inner excess field. Substituting Eq.~\eqref{eq:excess:potential:general} we obtain the following lenghty expression
\begin{align}
\tilde\varsigma_\pm &= 4\kappa^2 \frac{2\Xi - \breve\kappa \sinh(\tilde\zeta/4) (\Theta e^{\kappa\tilde y} - \Xi e^{-\kappa\tilde y}/\Theta(}{[2\breve\kappa\sinh(\tilde\zeta/4) + \Theta e^{\kappa\tilde y} - \Xi e^{-\kappa\tilde y}/\Theta]^2} \nonumber \\
&\mp 4\kappa^2 \frac{\kappa \sinh(\tilde\zeta/4) (\Theta e^{\kappa\tilde y} + \Xi e^{-\kappa\tilde y}/\Theta)}{[2\breve\kappa\sinh(\tilde\zeta/4) + \Theta e^{\kappa\tilde y} - \Xi e^{-\kappa\tilde y}/\Theta]^2},
\end{align}
where $\Xi$ and $\Theta$ are shorthands for $\Xi = (\kappa^2-\breve\kappa^2) \sinh^2(\tilde\zeta/4)$ and $\Theta = \kappa\cosh(\tilde\zeta/4) - \breve\kappa\sinh(\tilde\zeta/4)$. With this, the general form of (our approximation for) the ion distributions, uniformly valid in space, and in time from 	quasi-equilibrium, across the transition regime, to nonequilibrium, becomes
\begin{align}
c_\pm(x,t) =& \tilde\varsigma_\pm\left(\frac{1+x}{\epsilon},t\right) + \breve\varsigma_\pm\big(z_+(t),t\big) \nonumber\\
 &+ \hat c\left(\frac{1+x}{\sqrt{\epsilon}},t\right) - \bar c_o + \hat c\left(\frac{1-x}{\sqrt{\epsilon}},t\right) \nonumber\\
 &+ \breve\varsigma_\mp\big(z_-(t),t\big) + \tilde\varsigma_\mp\left(\frac{1-x}{\epsilon},t\right), \label{eq:uniformly:valid:gen}
\end{align}
where $z_\pm(t) = \big[|\bar J|(1 \pm x \mp y_o(t)) \pm \hat c_s(t)\big]/|\epsilon\bar J|^{2/3}$.

The resulting concentration profiles are shown in Fig.~\ref{fig:profiles120} for the solution at $V=120$, $\omega=0.3$, $\delta=0.3$, $\nu=0$, and $\epsilon=0.001$, displaying first injection of salt from the double layer into the diffusion layer, followed by re-uptake un the double layer, salt depletion in the diffusion layer with formation and growth of an extended space-charge region, and, finally, collapse of the space-charge layer when the cell current changes direction. The figure also compares the uniformly valid approximation to the full numerical solution from Fig.~\ref{fig:pnp120}, and shows very good agreement. The relative error, measured as $|c_\pm(y,t)-c_\pm^\mathrm{PNP}(y,t)|/\max_tc_\pm^\mathrm{PNP}(y,t)$, is below 4\% for all $y$ on the eight frames displayed in Fig.~\ref{fig:profiles120}.

%\begin{figure*}
%\includegraphics[width=0.85\textwidth]{compete.strongly.eps}
%\caption{Time average absolute potential drop across double layer.}
%\label{fig:time:average:potential}
%\end{figure*}

\begin{figure*}
\includegraphics[width=0.85\textwidth]{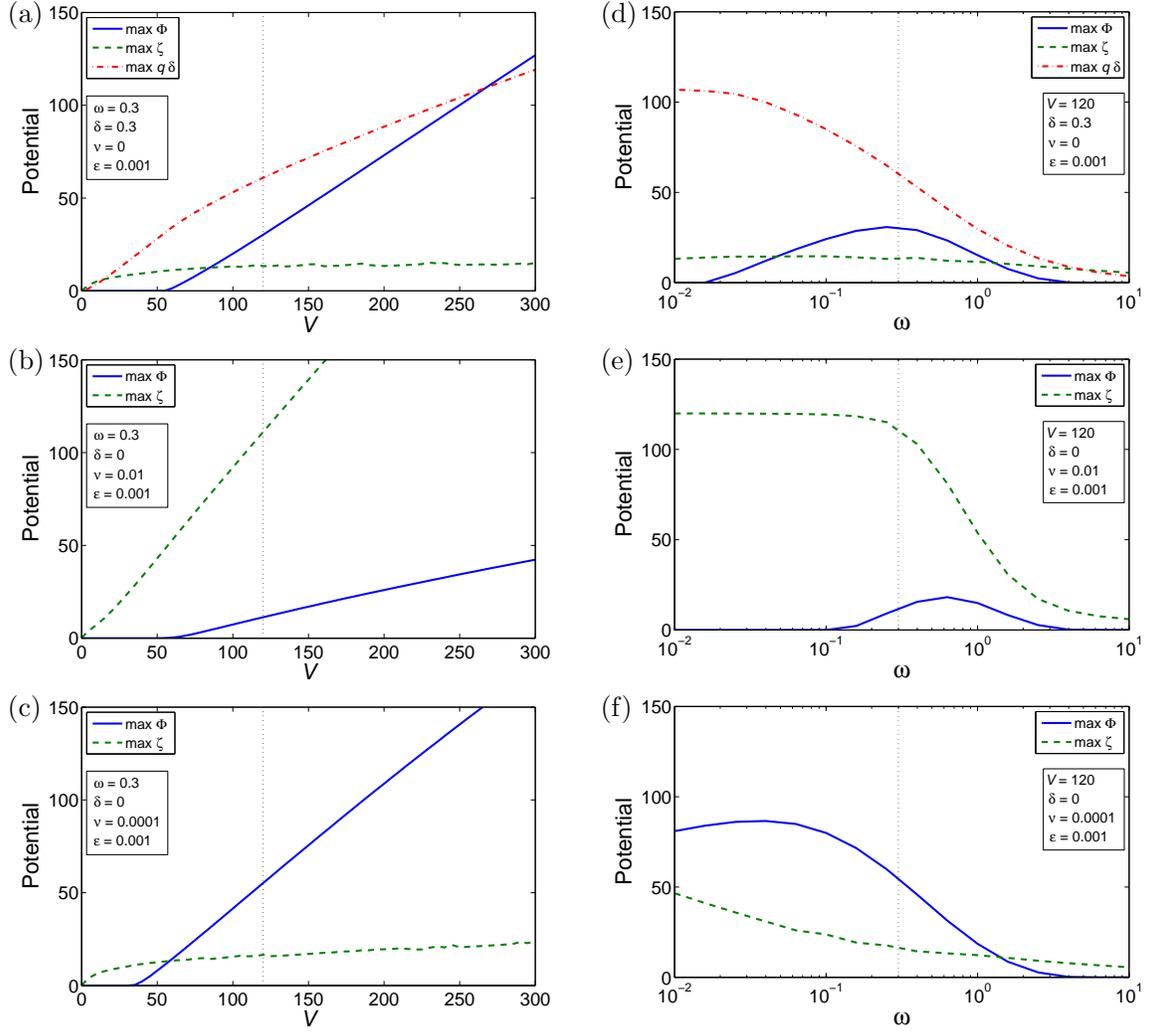}
\caption{(Color online) Peak voltages on space-charge layer, $\max_t\breve\Phi$ (solid), diffuse layer, $\max_t\tilde\zeta$ (dashed), and compact layer, $\max_t\tilde q\delta$ (dash-dot), for different values of the capacitance ratio $\delta$ and nominal ion volume fraction $\nu$. Panels (a) to (c) show results at $\omega = 0.3$ as a function of $V$, and (d) to (f) show results at $V = 120$ as a function of $\omega$. For the PB model, (a) and (d) show that the compact layer voltage $-\tilde q\delta$ dominates at $\delta = 0.3$, although $\breve\Phi$ becomes significant at large voltage. For the MPB model, (b) and (e) show that the diffuse-layer voltage $\tilde\zeta$ dominates at $\nu = 0.01$, while in (c) and (f) the space-charge layer-voltage $\breve\Phi$ dominates at $\nu = 0.0001$.}
\label{fig:time:average:potential}
%\label{fig:max:potential}
\end{figure*}

%%%%%%%%%%%%%%%%%%%%%%%%%%%%%%%%%%%%%%%%%%%%%%%%%%%%%%%%%%%%%%%%%%%%%%%%%%%%%%%
\subsection{Dominant voltage in double layer}

For electrochemical cells running with a dc Faradaic current, it is well known that concentration polarization can play a dominant role at large voltage, with the space-charge layer determining the overall current-voltage relation for the system~\cite{rubinstein1979,chu2005}. It is clear from Fig.~\ref{fig:potentials120} that although $\breve\Phi$ is smaller than the compact-layer voltage, $-\tilde q\delta$, it does affect the bulk current and slows down the charging process.

On the other hand, since $\breve\Phi$ depends on the capacitive current in ac, and since the diffuse-layer capacitance eventually drops when crowding starts to kick in, one might ask if the overall cell response will not be dominated by $\tilde\zeta$ at large voltage? Of course, that will depend on just how early the steric limit is reached, i.e., it depends on the nominal ion volume fraction $\nu$.

Figure~\ref{fig:potentials120nu} shows the strongly nonlinear response for the Bikerman model with two different values of $\nu$, at $V=120$, $\omega=0.3$, $\delta=0$, and $\epsilon=0.001$. In Fig.~\ref{fig:potentials120nu}(a) and (b) where $\nu = 0.01$, the diffuse-layer voltage $\tilde\zeta$ dominates in the cell while the space-charge layer voltage $\breve\Phi$ is negligible. Also the bulk ohmic potential drop $\bar J/\bar c_o$ is small because the driving frequency is below the characteristic frequency, $\omega<\omega_o$, cf. Fig.~\ref{fig:rc:time}. In Fig.~\ref{fig:potentials120nu}(c) and (d) where $\nu=0.0001$, the situation is the opposite: $\breve\Phi$ dominates over $\tilde\zeta$ in the non-equilibrium double layer, while $\bar J/\bar c_o$ dominates the overall cell response because this system is driven \emph{above} the characteristic frequency, $\omega>\omega_o$.

The competition between $\breve\Phi$, $\tilde\zeta$, and $-\tilde q\delta$ is investigated further in Fig.~\ref{fig:time:average:potential}, showing the peak values $\max_t\breve\Phi$, $\max_t\tilde\zeta$, and $\max_t\tilde q\delta$ as a function of driving voltage $V$ in panels (a) to (c) and frequency $\omega$ in panels (d) to (f). Figure~\ref{fig:time:average:potential}(a) and (d) shows results for the PB model with $\delta = 0.3$. Here the compact-layer voltage dominates, although $\breve\Phi$ grows to a significant fraction at large voltage. Note in Fig.~\ref{fig:time:average:potential}(d) that $\breve\Phi \propto \bar J_{{}}^{1/2}y_o^{3/2}$ peaks around the characteristic frequency $\omega_o \approx 0.3$: At higher frequencies the double layer is not fully charged so $\tilde w$ and $y_o$ decreases, while at lower frequencies it \emph{is} fully charged and hence $\bar J$ decreases.

Figure~\ref{fig:time:average:potential}(b) and (e) shows results for the MPB model with $\nu = 0.01$, where $\tilde\zeta$ completely dominates over $\breve\Phi$, whereas in Fig.~\ref{fig:time:average:potential}(c) and (f) with $\nu = 0.0001$, $\breve\Phi$ dominates over $\tilde\zeta$ at large voltage and not-so-high frequency.

The relative magnitude of the double-layer voltages can be understood from a simple estimate: Once steric effects dominate in the diffuse layer we have
\beq
\tilde\zeta = O(\nu\tilde q^2/2).
\eeq
For the space-charge layer we have $\bar J = O(\omega\tilde q)$ and $y_o = O(\epsilon\tilde w) = O(\epsilon\tilde q)$ so that
\beq
\breve\Phi = O(|\bar J|^{1/2}y_o^{3/2}/\epsilon) = O(\sqrt{\epsilon\omega}\tilde q^2).
\eeq
This is an important finding: With both $\tilde\zeta$ and $\breve\Phi$ scaling as $O(\tilde q^2)$ at large voltage, we expect $\tilde\zeta$ to dominate over $\breve\Phi$ for $\nu\gg\sqrt{\epsilon\omega}$, i.e., in systems with high nominal concentration (large $\nu$), large electrode separation (small $\epsilon$), and at low frequency ($\omega\ll\omega_o$). Conversely, we expect $\breve\Phi$ to play a dominant role at large voltage for systems with very dilute electrolytes and small (micro) electrode geometry, driven around the characteristic ($RC$) frequency.

%%%%%%%%%%%%%%%%%%%%%%%%%%%%%%%%%%%%%%%%%%%%%%%%%%%%%%%%%%%%%%%%%%%%%%%%%%%%%%%
\section{Summary and discussion\label{sec:conclusion}}

We have developed a dynamical model for the response of dilute electrolytes to large applied ac voltages, building on a body of theoretical work on diffuse-charge dynamics for both the weakly and strongly nonlinear regimes~\cite{bazant2004,chu2006,kilic2007b}, and on concentration polarization and space-charge layers in dc electrochemical systems running at steady-state conditions~\cite{rubinstein1979,rubinstein2001,chu2005}. Our original contributions are the solution in the oscillating diffusion layer, controlling the extent of the transient space-charge layer, and the uniformly valid formulation of the charge-voltage relation over the transition between quasi-equilibrium and non-equilibrium.

We have compared our asymptotic analysis for the PNP model to a full numerical solution of the PNP equations, and found good qualitative and quantitative agreement. The strongly nonlinear regime, characterized by strong concentration gradients in the diffusion layer, set in for $\sqrt{\epsilon\omega}\langle\tilde w\rangle = O(1)$, where the time-average excess salt concentration $\langle\tilde w\rangle$ depends on the driving frequency and voltage, but also on the intrinsic surface capacitance and crowding effects through $\delta$ and $\nu$, respectively. At very large voltage we argue that the cell response should be dominated by space-charge for $\nu\ll\sqrt{\epsilon\omega}$ and by crowding effects for $\nu\gg\sqrt{\epsilon\omega}$.

Recently, Beunis \emph{et al.}~\cite{beunis2008} presented an analysis of the \emph{transient} response to a dc step large enough to introduce transient space-charge. They analyse four extreme cases: The ``double-layer limited" ($V\ll1$, $\epsilon\ll1$), ``diffusion limited" ($V\ll1$, $\epsilon\gg1$), ``geometry limited" ($V\gg1$, $\epsilon\gg1/\sqrt{V}$), and ``space-charge limited" ($V\gg1$, $\epsilon\ll1/\sqrt{V}$), and develop closed form analytical solutions in each of those limits. In particular, for the space-charge limited response, setting $\breve\Phi = V$ and $y_o = \epsilon|\tilde q|$, they predict a characteristic $O(t^{-3/4})$ dependence in the bulk current, which they verify by experiments on a system with surfactant micelle charge carriers.

While the simple analytical results of Ref.~\cite{beunis2008} provides important insight to the \emph{limiting} case when the space-charge layer completely dominates the response, our dynamical model is \emph{uniformly} valid from small to very large voltage. The general applicability, however, comes at the expense that our strongly nonlinear model requires a set of integro-differential-algebraic equations to be solved numerically. In comparison, the weakly nonlinear ``circuit" model, which neglects any perturbations to the bulk and diffusion-layer electrolyte concentration at leading order, can be formulated as a simple ordinary differential equation.

We also consider steric effects of finite-sized ions in blocking cells under large ac voltages.  This leads to a novel and strong dependence on the bulk volume fraction of ions $\nu$, which acts as a third dimensionless parameter, along with $\epsilon$ and $V$, to determine different dynamical regimes. We consider the regime of ionic liquids $\nu = O(1)$, up to the molten salt limit $\nu \approx 1$, and find that strongly nonlinear regime disappears with increasing $\nu$ due to dominant  steric effects, which prevent the double layers from adsorbing significant numbers of ions from the bulk. The classical diffuse layer is effectively replaced by a molecular condensed layer.  As a result, we justify the use of weakly nonlinear circuit models, as in Refs.~\cite{kornyshev2007,federov2008,federov2008b} to describe the dynamics of ionic liquids up to very large, time-dependent applied voltages.

In many cases,  even in dilute electrolytes, the nonlinear circuit model may actually give good account for \emph{overall} cell current-voltage response, in particular when the double-layer capacitance is dominated by the compact layer, or when crowding effects set in and a condensed layer forms at the electrode. However, this does not necessarily mean that the weakly nonlinear analysis will account well for \emph{all} aspects of the electrokinetic response, such as ac-electroosmotic fluid motion and pumping.

%%%%%%%%%%%%%%%%%%%%%%%%%%%%%%%%%%%%%%%%%%%%%%%%%%%%%%%%%%%%%%%%%%%%%%%%%%%%%%%
\subsection{Two or more dimensions}

The boundary-layer analysis can be easily extended to higher dimensions, provided the electrode geometry is smooth enough to be considered locally flat on the boundary-layer length scale. Then, the steady-state bulk response becomes
\beq
\nablabf\cdot\bar\Jbf = 0 \quad \mbox{ and } \quad \nablabf\cdot\bar\Fbf = 0,
\eeq
where $\bar\Jbf = -\bar c\nablabf\bar\phi$ is the current, $\bar\Fbf = -\nablabf\bar c + Pe\langle\bar\ubf\rangle\bar c$ is the salt flux, and $\bar c = \bar c(\textbf{r})$ is constant in time but not in space. The last term in the flux describes advection by the average fluid velocity $\langle\bar\ubf\rangle$; $Pe = u_0L/D$ is the P\'eclet number, where $u_0 = \varepsilon(kT/ze)^2/\eta L$ is the electroosmotic (EO) velocity scale, and $\eta$ is the dynamic viscosity. In the surface conservation laws, tangential flux through the highly charged diffuse layer must be taken into account~\cite{bikerman1933,deryagin1969,dukhin1993,chu2006,chu2007}, leading to
\begin{align}
\partial_t\tilde q &= \nbf\cdot\bar\Jbf + \epsilon\nablabf_s\cdot\tilde\Jbf_s, \nonumber \\
\partial_t\tilde w &= \tilde F_o + \epsilon\nablabf_s\cdot\tilde\Fbf_s, \label{eq:surface:conservation:tangent}
\end{align}
where $\nablabf_s$ is the tangential gradient, and $\tilde\Jbf_s$ and $\tilde\Fbf_s$ are the surface excess current and salt flux, respectively, due to surface migration and EO convection. For the PB model it can be shown that
\begin{align}
\tilde\Jbf_s &= (1+Pe)(\tilde w\nablabf_s\bar\phi + \tilde q\nablabf_s\log\hat c_s), \nonumber \\ %\label{eq:surface:current} \\
\tilde\Fbf_s &= (1+Pe)(\tilde q\nablabf_s\bar\phi + \tilde w\nablabf_s\log\hat c_s). \label{eq:surface:flux}
\end{align}
The same results also apply for the MPB model, provided the ion mobility in the highly crowded double layer is equal to that in the bulk, which is, however, questionable~\cite{dukhin1993,large}.

Assuming transverse convection is weak enough, $\epsilon Pe |\bar\ubf|\ll1$, the diffusion layer can still be modelled by simple 1D diffusion in the normal direction, with the concentration given by Eq.~\eqref{eq:convolution:noneq}. Matching with the steady solution in the bulk is then obtained by
\beq[salt:flux:bc:2D]
\nbf\cdot\bar\Fbf = \langle\tilde F_o\rangle = \epsilon\langle\nablabf_s\cdot\tilde\Fbf_s\rangle,
\eeq
since the oscillating diffusion layer does not accumulate any salt on time average, but only acts as a buffer zone for the periodic flux in and out of the diffuse layer. In this way, surface conduction can drive bulk concentration gradients even in the steady-state response~\cite{chu2006}.

The bulk fluid motion is driven primarily by EO slip from the boundary layers. For the quasi-equilibrium double layer, the effective tangential slip velocity according to PB theory becomes
\beq
\bar\ubf_s = \tilde\zeta\nablabf_s\bar\phi + 4\log(\cosh(\tilde\zeta/4))\nablabf_s\log\hat c_s.
\eeq
For induced-charge electroosmosis (ICEO) where both $\tilde\zeta$ and $\bar\phi$ depend on the external driving voltage, the velocity scales as $O(V^2)$ at low voltage~\cite{gonzalez2000,iceo2004b}. At larger voltage, PB theory predicts stall of $\tilde\zeta$ and scaling only as $O(V\log V)$~\cite{olesen2006}, whereas the MPB model predicts a return to the $O(V^2)$ scaling once crowding effects set in~\cite{storey2008}. Or, this assumes the viscosity in the highly crowded double layer is equal to the bulk value; if it is significantly reduced one might expect the velocity to scale as $O(V\log\sqrt{2/\nu})$, where $\log\sqrt{2/\nu}$ is the diffuse layer voltage in the dilute part \emph{outside} the condensed layer~\cite{large}.

When the double layers are driven out of quasi-equilibrium, they are still governed by surface conservation laws like Eq.~\eqref{eq:surface:conservation:tangent}. The surface fluxes $\tilde\Jbf_s$ and $\tilde\Fbf_s$ are no longer given by Eq.~\eqref{eq:surface:flux}, but they remain dominated by the inner diffuse layer, since the concentration (and hence conductivity) in the space-charge layer is low.

If the space-charge layer does not contribute much to the surface fluxes, it plays a major role on EO fluid motion: The voltage $\breve\Phi$ drives a Smoluchowski-type slip velocity
\beq
\bar\ubf_s = \breve\Phi\nablabf_s\bar\phi,
\eeq
for which Dukhin and co-workers coined the term ``electroosmosis of the second kind" (EO2) to distinguish it from the quasi-equilibrium response~\cite{dukhin1991}. This phenomenon has been studied extensively in the context of nonlinear electrophoresis of conductive particles made from ion-exchanger material~\cite{dukhin1991,ben2004}.

Since many ac electrokinetic experiments involve microelectrodes and applied voltages of a few volt~\cite{ramos1998,green2000b,levitan2005,fagan2005,wang2004}, including investigations on ac electroosmotic micropumps~\cite{brown2000,studer2004,gregersen2007}, and since our analysis has shown that this is enough to create strong concentration polarization and transient space-charge around the electrodes, we believe that EO2 could be important for interpreting the experimental results.

Rubinstein and Zaltzmann showed that EO2 renders linearly unstable the quiescent solution of concentration polarization on a planar permselective membrane running at dc, leading to spontaneous formation of vortex pairs that stir up the concentration profile in the diffusion layer, which in turns enables the passage of ``super-limiting" current through the membrane~\cite{rubinstein2001,rubinstein2005}. Presumably, a similar instability could occur for transient space-charge layers, although the threshold voltage may depend on whether the space-charge layer voltage $\breve\Phi$ is dominating in the overall cell response or not.

%%%%%%%%%%%%%%%%%%%%%%%%%%%%%%%%%%%%%%%%%%%%%%%%%%%%%%%%%%%%%%%%%%%%%%%%%%%%%%%
\subsection{Diffusive dynamics in the bulk}

The transient response in the bulk, while the cell relaxes towards the
steady-state periodic solution, or as arising from a slowly varying ac
voltage amplitude $V_\mathrm{ext} = V(\bar t)\sin(\omega t) + V_o(\bar t)$,
is governed by diffusive dynamics on the slow time scale $\bar t = \epsilon
t$ \beq
\partial_{\bar t}\bar c = -\nablabf\cdot\bar\Fbf = \nablabf^2\bar c - Pe\langle\bar\ubf\rangle\cdot\nablabf\bar c,
\eeq
driven by the time-average flux into the double layer
\beq
\nbf\cdot\bar\Fbf = \epsilon\langle\nablabf_s\cdot\tilde\Fbf_s\rangle - \partial_{\bar t}\langle\tilde w\rangle.
\eeq
Here the time averages are taken on the $RC$ time scale $t$, i.e., for each timestep taken on the slow time scale, we require the periodic response of the boundary layers on the $RC$ time scale to be determined.

%%%%%%%%%%%%%%%%%%%%%%%%%%%%%%%%%%%%%%%%%%%%%%%%%%%%%%%%%%%%%%%%%%%%%%%%%%%%%%%
\subsection{Faradaic reactions and general electrolytes}

Much interest on electrokinetics is of course associated with electrochemistry and reactions on electrodes that are not blocking but support the passage of a Faradaic current. Then the surface conservation laws are enriched (is this the right word??) by the injection of a Faradaic current $J_\mathrm{ext}$ at the electrode surface, and an associated salt flux $F_\mathrm{ext}$.

Depending on the charge-transfer resistance and on the driving frequency and voltage, the Faradaic current may be small compared to the capacitive current in ac, or it may completely dominate the charging dynamics of the double layer. For reactions controlled by Butler-Volmer kinetics, where the reaction rate grows exponentially with (compact-layer) voltage, the latter should be the case at sufficiently large voltage. Of course, this implies that the $RC$ time scale, formed by the bulk ohmic resistance and double-layer capacitance, may not be appropriate for describing the system response.

Moreover, for many reactions it is not sufficient to assume a binary electrolyte since neutral reaction products also play an important role. The general problem of the dynamic response for general electrolytes seems daunting, and is a challenge even for the steady-state response~\cite{urtenov2007}.

\acknowledgments
This work was supported in part by the U.S. National Science Foundation under contract DMS-0707641 (MZB).  The authors thank ESPCI for hospitality during our collaboration.

\appendix*

%%%%%%%%%%%%%%%%%%%%%%%%%%%%%%%%%%%%%%%%%%%%%%%%%%%%%%%%%%%%%%%%%%%%%%%%%%%%%%%
\section{\label{sec:painleve}}

\begin{figure*}
\includegraphics[width=0.9\textwidth]{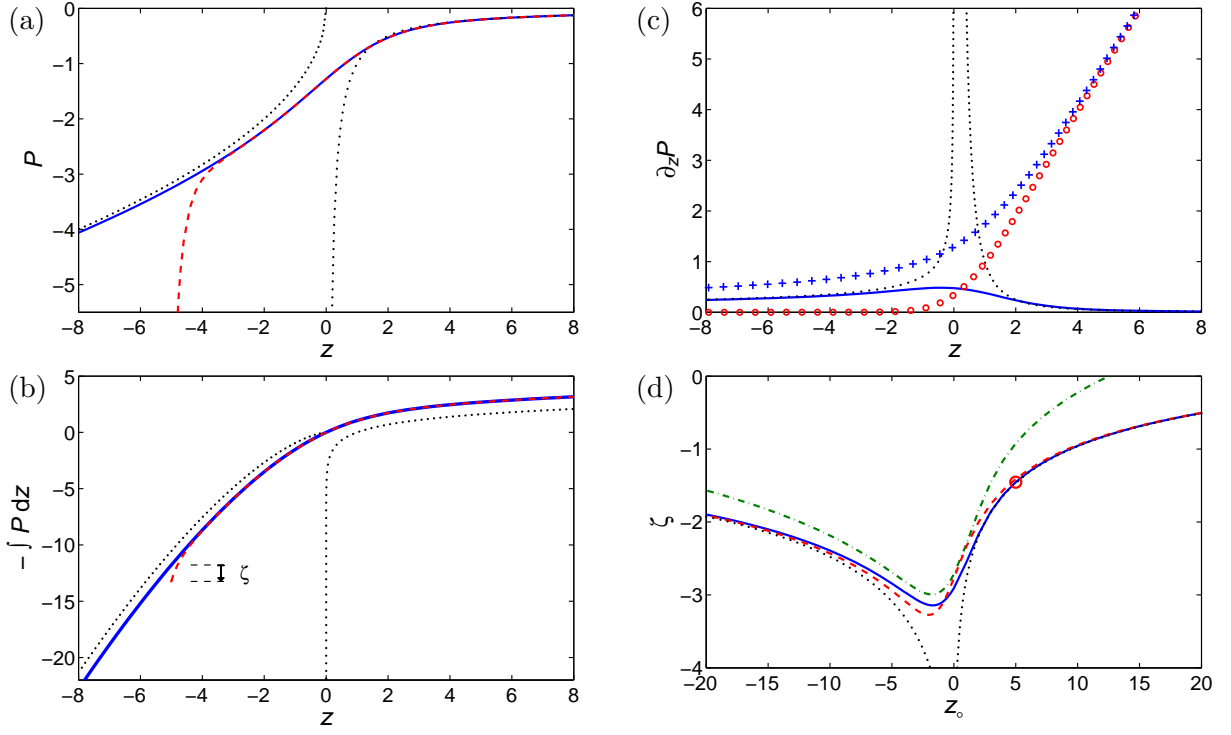}
\caption{(Color online) Solution in ``Smyrl-Newman" transition layer in terms of Painlev\'e transcendents. (a) Rescaled electric field $P(z)$ (solid) and leading order terms from asymptotic expansion in space-charge layer, $z\ll-1$, and diffusion layer, $z\gg1$, (dotted). The dashed line shows a solution $\tilde P$ to Eq.~\eqref{eq:painleve} with b.c. $\tilde P = \tilde P_o = -10$ applied at $z = -z_o = -5$. (b) Potential variation $\phi = -\int_0^z P(z')dz'$ in transition layer (solid), and leading order terms from asymptotic approximation (dotted). Again, the dashed line shows the result for the solution $\tilde P$. (c) Rescaled charge distribution $\partial_zP$ (solid) and leading order terms from asymptotic approximation (dotted). Symbols show individual ion concentrations, cf. Eq.~\eqref{eq:rescaled:concentration}. (d) Excess voltage $\tilde\zeta$ as a function of $z_o$ with $\tilde P_o = -10$ (solid), and leading order terms from asymptotic approximation (dotted). The circle marks $z_o=5$ corresponding to $\tilde P$ from panels (a) and (b), and the dashed and dash-dot lines show the approximation by Eqs.~\eqref{eq:voltage:charge:painleve} and \eqref{eq:voltage:charge:zaltzman}, respectively.}
\label{fig:painleve}
\end{figure*}

The theory for non-equilibrium double layers was originally developed for electrochemical systems passing a dc current in steady state~\cite{grafov1962,chernenko1963,smyrl1967}, with boundary conditions representing either normal flux of ions into a permeable electrodialysis membrane~\cite{rubinstein1979} or via Faradaic charge-transfer reactions at an electrode~\cite{chu2005}. If we consider a cationic space-charge layer formed with negative voltage on the left electrode, the Nernst--Planck equations become
\begin{align}
F_+ &= -\partial_yc_+ - c_+\partial_y\phi = 2\bar J < 0, \label{eq:smyrl:newman:flux1} \\
F_- &= -\partial_yc_- + c_-\partial_y\phi = 0. \label{eq:smyrl:newman:flux2}
\end{align}
For a dc electrochemical system, this holds across the entire cell. For an ac system driven at large voltage, this holds in the inner diffuse, space-charge, and ``Smyrl-Newman" transition layers because the charging process is dominated by uptake of counterions (cations) rather than expulsion of coions (anions), such that $|F_+| \gg |F_-|$. However, in the bulk region and in the diffusion layer we have primarily ohmic transport and $F_+ \approx -F_- \approx \bar J$. After some manipulations on Eqs.~\eqref{eq:smyrl:newman:flux1}, \eqref{eq:smyrl:newman:flux2}, and \eqref{eq:poisson} the problem is reduced to a single master equation for the electric field~\cite{chu2005}
\beq
\varepsilon^2\partial_y^2E - \frac{1}{2}\varepsilon^2E^3 - |\bar J|(y-y_o^*)E = |\bar J|,
\eeq
where $E = -\partial_y\phi$ and $y_o^*$ is an integration constant. For $y_o^*>0$ it is equal to the width $y_o$ of the space-charge layer, whereas for $y_o^*<0$ we interpret it as $y_o^* = -\hat c_s/|J|$~\cite{chu2005}. Rescaling with
\beq
E = \frac{|\bar J|^{1/3}}{\varepsilon^{2/3}}\, P \quad\mbox{and}\quad y-y_o^* = \frac{\varepsilon^{2/3}}{|\bar J|^{1/3}}\, z
\eeq
we then arrive at
\beq[painleve]
\partial_z^2P = \frac{1}{2}P^3 + zP + 1
\eeq
This is an instance of the second order ordinary differential equation with Painlev\'e property (i.e., all movable singularities are poles) defining the Painlev\'e transcendents of the second kind. The connection between steady dc current in electrochemical cells and Painlev\'e transcendents was first noted by Grafov and Chernenko~\cite{grafov1962,chernenko1963}. Eq.~\eqref{eq:painleve} has a unique ``transition layer" solution $P(z)$ with no poles on the real axis and the following asymptotic behaviour
\beq[smyrl:newman:asymptotics]
P(z) =
\left\{\begin{array}{rl}
-1/z + O(1/z^4) & \mbox{for } z\to+\infty, \\
-\sqrt{-2z} + O(1/z) & \mbox{for } z\to-\infty.
\end{array}\right.
\eeq
The detailed shape of $P(z)$ is shown in Fig.~\ref{fig:painleve}(a) and compared with the leading order asymptotics.
Fig.~\ref{fig:painleve}(b) shows the potential variation
\beq
\phi(z) = -\int_0^z P(z')\id z',
\eeq
and Fig.~\ref{fig:painleve}(c) displays the rescaled charge density
\beq
\frac{\rho}{|\epsilon\bar J|^{2/3}} = \partial_zP,
\eeq
and individual ion concentrations
\beq[rescaled:concentration]
\frac{c_\pm}{|\epsilon\bar J|^{2/3}} = z + \frac{1}{2}P^2 \pm \partial_zP.
\eeq

\subsection*{Boundary layer}

The form of $P(z)$ describes the solution in the interior of the electrochemical cell. However, $P(z)$ generally does not satisfy the boundary conditions at the electrodes confining the cell. Imposing b.c.'s on the solution gives rise to boundary layers, that can be understood mathematically as originating from poles in the solution, located outside the domain of the physical cell.

We focus on the behaviour at the left electrode, and consider a solution $\tilde P(z)$ to Eq.~\eqref{eq:painleve} on the interval $z\in[-z_o,\infty)$, with boundary conditions $\tilde P(-z_o) = \tilde P_o$ and $\tilde P(\infty) = 0$. Here
\beq
-z_o = \frac{\hat c_s - |\bar J|y_o}{|\epsilon\bar J|^{2/3}}
\eeq
corresponds to the rescaled position of the electrode. Figure~\ref{fig:painleve}(a) shows the result for $z_o = 5$ and $\tilde P_o = -10$: The excess field is rapidly screened out, and for $z \gtrsim -4$ we see that $\tilde P(z)$ follows $P(z)$ closely.

Let us introduce the \emph{excess} field $\tilde R = \tilde P - P$ in the boundary layer. Substituting into Eq.~\eqref{eq:painleve} we obtain
\beq[excess:field:1]
\partial_z^2\tilde R = \frac{1}{2}\tilde R^3 + \frac{3}{2}P\tilde R^2 + k^2\tilde R,
\eeq
where $k = \sqrt{3P^2/2+z}$. Since the boundary layer is thin, it is reasonable to approximate the variable coefficients with constants $P_o = P(-z_o)$ and $k_o = \sqrt{3P_o^2/2-z_o}$ to get
\beq[excess:field:2]
\partial_z^2\tilde R = \frac{1}{2}\tilde R^3 + \frac{3}{2}P_o\tilde R^2 + k_o^2\tilde R.
\eeq
This approximation is crudest for $z_o$ close to zero, where the local screening length $1/k_o$ has a maximum; for $z_o\ll-1$ and $z_o\gg1$ we have $k_o\approx\sqrt{-z_o}$ and $k_o\approx-P_o\approx\sqrt{2z_o}$, respectively.

Integrating twice on Eq.~\eqref{eq:excess:field:2} we obtain
\beq
\partial_z\tilde R = -\tilde R\sqrt{\tilde R^2/4 + P_o\tilde R + k_o^2},
\eeq
and
\begin{align}
z+z_o = \frac{1}{k_o}\Bigg[&\sinh^{-1}\Bigg(\frac{P_o\tilde R+2k_o^2}{|\tilde R|\sqrt{P_o^2/2-z_o}}\Bigg) \nonumber \\
& - \sinh^{-1}\Bigg(\frac{P_o\tilde R_o+2k_o^2}{|\tilde R_o|\sqrt{P_o^2/2-z_o}}\Bigg)\Bigg],
\end{align}
from which
\beq
\tilde R = -\frac{2k_o^2}{P_o-\mathrm{sign}(\tilde R_o)\sqrt{P_o^2/2-z_o}\sinh\big[k_o(z+z^*)\big]},
\eeq
where
\beq
z^* = z_o + \frac{1}{k_o}\sinh^{-1}\bigg(\frac{P_o\tilde R_o+2k_o^2}{|\tilde R_o|\sqrt{P_o^2/2-z_o}}\bigg) .
\eeq
Finally, the excess potential $\tilde\psi = \tilde\phi - \phi$ is found by integrating $\tilde R = -\partial_z\tilde\psi$ to
%\beq
%\tilde\psi = 4\tanh^{-1}\left[\frac{P_o\tanh[k(z+z^*)/2]-\mathrm{sign}(\tilde R_o)\sqrt{P_o^2/2-z_o}}{k}\right] - 4\tanh^{-1}\left[\frac{P_o-\mathrm{sign}(\tilde R_o)\sqrt{P_o^2/2-z_o}}{k}\right]
%\eeq
\begin{align}
\tilde\psi &= 4\tanh^{-1}\Bigg[\frac{\tanh(\tilde\zeta/4)e^{-k_o(z+z_o)}}{1+\frac{P_o}{k_o}\tanh(\tilde\zeta/4)[1-e^{-k_o(z+z_o)}]}\Bigg] \\
 &= 2\log\Bigg[\frac{1 + \frac{k_o+P_o}{k_o-P_o}e^{\tilde\zeta/2} + (1-e^{\tilde\zeta/2})e^{-k_o(z+z_o)}}{1+\frac{k_o+P_o}{k_o-P_o}\big[e^{\tilde\zeta/2}+(1-e^{\tilde\zeta/2})e^{-k_o(z+z_o)}\big]}\Bigg]
\end{align}
Here $\tilde\zeta = \tilde\psi(-z_o)$ is determined through the boundary condition $\tilde P_o = P_o + \tilde R_o$ as
%\beq[charge:voltage:painleve]
%\tilde P_o = P_o + 2k_o\sinh(\tilde\zeta/2)\big[1+P_o\tanh(\tilde\zeta/4)/k_o\big],
%\eeq
%or
\begin{align}
\tilde\zeta &= -2\log\Bigg(\sqrt{\bigg[\frac{\tilde P_o+P_o}{2(k_o-P_o)}\bigg]^2+\frac{k_o+P_o}{k_o-P_o}}-\frac{\tilde P_o+P_o}{2(k_o-P_o)}\Bigg) \label{eq:voltage:charge:painleve} \\
 &= 2\sinh^{-1}\Bigg(\frac{\tilde P_o + P_o}{2\sqrt{k_o^2-P_o^2}}\Bigg) + \log\bigg(\frac{k_o-P_o}{k_o+P_o}\bigg)
\end{align}
Using $k_o\approx \sqrt{-z_o}$ and $P_o\approx 1/z_o$ for $z_o\ll-1$ it is easily verified that Eq.~\eqref{eq:voltage:charge:painleve} reduces to Chapman's formula
\beq[chapmanz]
\tilde\zeta \approx 2\sinh^{-1}\bigg(\frac{\tilde P_o}{2\sqrt{-z_o}}\bigg)
\eeq
in quasi-equilibrium, and, similarly, using $k_o\approx -P_o\approx \sqrt{2z_o}$ for $z_o\gg1$ we find
\beq[voltage:charge:dilutez]
\tilde\zeta \approx -2\log\bigg(\frac{1}{2}-\frac{\tilde P_o}{\sqrt{8z_o}}\bigg),
\eeq
in accordance with Eq.~\eqref{eq:voltage:charge:dilute} in non-equilibrium. However, while Eqs.~\eqref{eq:chapmanz} and \eqref{eq:voltage:charge:dilutez} diverge for $z_o\to0$ at fixed $\tilde P_o$, our general result \eqref{eq:voltage:charge:painleve} only displays a local maximum. This is shown in Fig.~\ref{fig:painleve}(d) where $\tilde\zeta$ is plotted as function of $z_o$ for $\tilde P_o = -10$. The figure also compares our result to the excess voltage from a direct numerical solution for $\tilde P$, and it is seen that Eq.~\eqref{eq:voltage:charge:painleve} slightly overestimates $\tilde\zeta$ for $z_o<0$ and underestimates it for $z_o>0$.

\subsubsection*{Singular transcendents}

Zaltzman and Rubinstein \cite{zaltzman2007} systematically studied the transition from quasi-equilibrium to nonequilibrium by considering various ranges for the parameter $z_o$, solving appropriate approximations to the Painlev\'e equation in each range. In the transition regime they approach the innermost part of the inner diffuse layer by an algebraically decaying solution, matched to a \emph{singular} solution of the full original Painlev\'e equation~\eqref{eq:painleve} in the outer part, writing
\beq
\tilde P = -\frac{2}{z+z_0-2/\tilde P_o} + \frac{2}{z+z_o} + P^{\dagger}(z;z_o).
\eeq
Here $2/(z+z_o) + P^\dagger(z;z_o)$ is the \emph{regular} part of a solution $P^\dagger(z;z_o)$ with a simple pole at $-z_o$, $P^\dagger(z;z_o)\sim-2/(z+z_o)$ for $z\to-z_o$, and $P^\dagger(\infty;z_o) = 0$. This allows them to write
\beq[voltage:charge:zaltzman]
\tilde\zeta = -2\log(-\tilde P_o) + \varphi(z_o)
\eeq
where
\begin{align}
\varphi(z_o) = \lim_{z\to\infty}\bigg[&\int_{z_o}^zP^\dagger(z';z_o) + \frac{2}{z'+z_o} - P(z')\,\id z' \nonumber \\
 & - \log(z+z_o)\bigg].
\end{align}
Their approximation is highly accurate for $z_o$ in the transition range and large enough $\tilde P_o\ll-1$, and it is simple to evaluate once the function $\varphi(z_o)$ has been tabulated.
For comparison, our result \eqref{eq:voltage:charge:painleve} has a finite error for $z_o$ in the transition range, an error that does not vanish at large $\tilde P_o$ but tends to a finite value, essentially being due to our approximation of the variable screening ``constant" $k(z)$ in the tail of the excess field by a real constant $k_o$. On the other hand, our result matches fully with the quasi-equilibrium and nonequilibrium limits, allowing us to use a single formulation of the charge-voltage relation for the entire dynamic solution procedure.

%%%%%%%%%%%%%%%%%%%%%%%%%%%%%%%%%%%%%%%%%%%%%%%%%%%%%%%%%%%%%%%%%%%%%%%%%%%%%%%

%\bibliography{elec19}

\end{document}